\documentclass[preprint,12pt,authoryear]{elsarticle}
\usepackage{setspace}
\usepackage[margin=2.54cm]{geometry}
\usepackage{times}
\usepackage{stmaryrd} 
\usepackage[ansinew]{inputenc}
\usepackage{csquotes} 
\usepackage{titlesec, titleps, titletoc} 
\usepackage[normalem]{ulem} 
\usepackage{units} 
\usepackage{rotating}
\usepackage{nicefrac}
\usepackage{icomma}
\usepackage{natbib} 
\usepackage{scrpage2} 
\usepackage{listings} 
\usepackage{fullpage,enumerate, enumitem}
\usepackage{url} 
\usepackage{float} 
\usepackage[colorlinks]{hyperref} 
\hypersetup{citecolor=blue}
\usepackage{amsmath, amssymb, amsfonts, amsthm,psfrag,graphicx,color,stmaryrd,mathrsfs,MnSymbol}
\usepackage{multicol, blindtext}
\usepackage{ragged2e, array} 
\usepackage{longtable} 
\usepackage{subfig}
\usepackage{bm}

\usepackage{stmaryrd} 
\usepackage{csquotes} 
\usepackage{titlesec, titleps, titletoc} 
\usepackage[normalem]{ulem} 
\usepackage{units} 
\usepackage{rotating}
\usepackage{nicefrac}
\usepackage{icomma}
\usepackage{natbib} 
\usepackage{scrpage2} 
\usepackage{listings} 
\usepackage{fullpage,enumerate, enumitem}
\usepackage[ansinew]{inputenc}
\usepackage{url} 
\usepackage{float} 
\usepackage[colorlinks]{hyperref} 
\hypersetup{citecolor=blue}
\usepackage{amsmath, amssymb, amsfonts, amsthm,psfrag,graphicx,color,stmaryrd,mathrsfs,MnSymbol}
\usepackage{multicol, blindtext}
\usepackage{ragged2e, array} 
\usepackage{longtable} 
\usepackage{subfig}
\usepackage{bm}

\newtheorem{theorem}{Theorem}

\newtheorem{definition}[theorem]{Definition}

\usepackage[T1]{fontenc}

\swapnumbers
\theoremstyle{plain}

\theoremstyle{definition}

\theoremstyle{remark}

\usepackage{setspace}

\DeclareMathAlphabet{\mathpzc}{OT1}{pzc}{m}{it}

\title{\textsc{Regime switching vine copula models for global equity and volatility indices}}

\author{$\text{\textsc{holger fink}}^{a,1}\text{\textsc{, yulia klimova}}^{b,2}\text{\textsc{, claudia czado}}^{b,3}\text{\textsc{, and jakob stöber}}^{b,4}$
}

\date{\today}

\begin{document}

\maketitle
\begin{multicols}{2}[]
\begin{center}
\textsc{\noindent\hrulefill\ abstract \noindent\hrulefill}
\end{center}
\small{For nearly every major stock market there exist equity and implied volatility indices. These play important roles within finance: be it as a benchmark, a measure of general uncertainty or a way of investing or hedging. It is well known in the academic literature, that correlations and higher moments between different indices tend to vary in time. However, to the best of our knowledge, no one has yet considered a global setup including both, equity and implied volatility indices of various continents, and allowing for a changing dependence structure. We aim to close this gap by applying Markov-switching $R$-vine models to investigate the existence of different, global dependence regimes. In particular, we identify times of \enquote{normal} and \enquote{abnormal} states within a data set consisting of North-American, European and Asian indices. Our results confirm the existence of joint points in time at which global regime switching takes place.}
\columnbreak
\begin{center}
\textsc{\noindent\hrulefill\ authors info \noindent\hrulefill}
\end{center}
\small{$^a$Institute of Statistics, Ludwig-Maximilians-Universit\"at M\"unchen, Akademiestr. 1/I, 80799 Munich, Germany\\
$^b$Department of Mathematics, Technische Universität München, Boltzmannstra\ss e 3, 85748 Garching Germany\\
$^1$holger.fink@stat.uni-muenchen.de\\
$^2$yulia.klimova@gmx.de\\
$^3$cczado@ma.tum.de\\
$^4$stoeber@ma.tum.de\\
}
\begin{center}
\textsc{\noindent\hrulefill\ keywords \noindent\hrulefill}
\end{center}
\small{regular vine copulas, Markov switching, implied volatility index, equity index, global dependence regimes}
\begin{center}
\textsc{\noindent\hrulefill\ jel subject classification \noindent\hrulefill}
\end{center}
\small{C58, C52, C10, G10}
\end{multicols}
\noindent\hrulefill

\section{Introduction}\label{sec1}
Based on the early work of \cite{Brenner1989} and \cite{Whaley93}, the Chicago Board Options Exchange (CBOE) started calculating its implied volatility index VIX, today well-known as a 'fear gauge', back in 1993. Thereby, they introduced not only a new asset class but also the first general index for market uncertainty. Early academic work therefore focussed mainly on the relationship between the VIX and realized volatility (cf. \cite{fleming1995}, \cite{fleming1998} or \cite{christensen1998}) and on pricing volatility derivatives (cf. \cite{Gruen96}).

Over the years, other global exchanges followed suit leading to today's availability of volatility indices for nearly every major equity gauge. Accordingly, practical applications and academic research have broadened as well. For once, volatility indices can provide additional information for pricing equity options (cf. \cite{Hao12} or \cite{KLY14}). From a portfolio management perspective, similar to the well-known leverage effect initially discussed by \cite{black1976}, especially the usually observed (asymmetric) negative dependence between implied volatility and equity indices can be useful for hedging purposes and risk management (cf. the recent work of \cite{Allenetal13} based on quantile regression or \cite{NXW15} on asymmetric volatility clustering). Additionally, implied volatility is in discussion to potentially have some predictive power on future returns as well (cf. \cite{Banerjee07} or \cite{Rubb14}) even though such results may not be generally valid as recently shown by \cite{ES15} and therein for currency carry strategies or \cite{MOS15} for fixed income markets. Finally, all these potential applications justify research on the modeling and prediction of volatility indices themselves as well (cf. \cite{DPS07} or \cite{Fernandes14}).

In this paper, however, we will contribute to a more macro-type application. As shown by the financial crisis back in 2007/2008 or the very recent sell-off in global equity markets in early 2016, major shifts in market sentiment are often not restricted to some area of the world but seem to quickly spread globally. Various researcher (as \cite{A07}, \cite{peng2012}, \cite{ACF13}, \cite{JM14}, \cite{Nomikos13} or \cite{SWe15}) discussed in the past that, for such structural breaks within the dependencies between equity markets, (implied) volatility might play an important role, as well.

\begin{figure}[htbp]
\centering
\includegraphics[height=10.5cm]{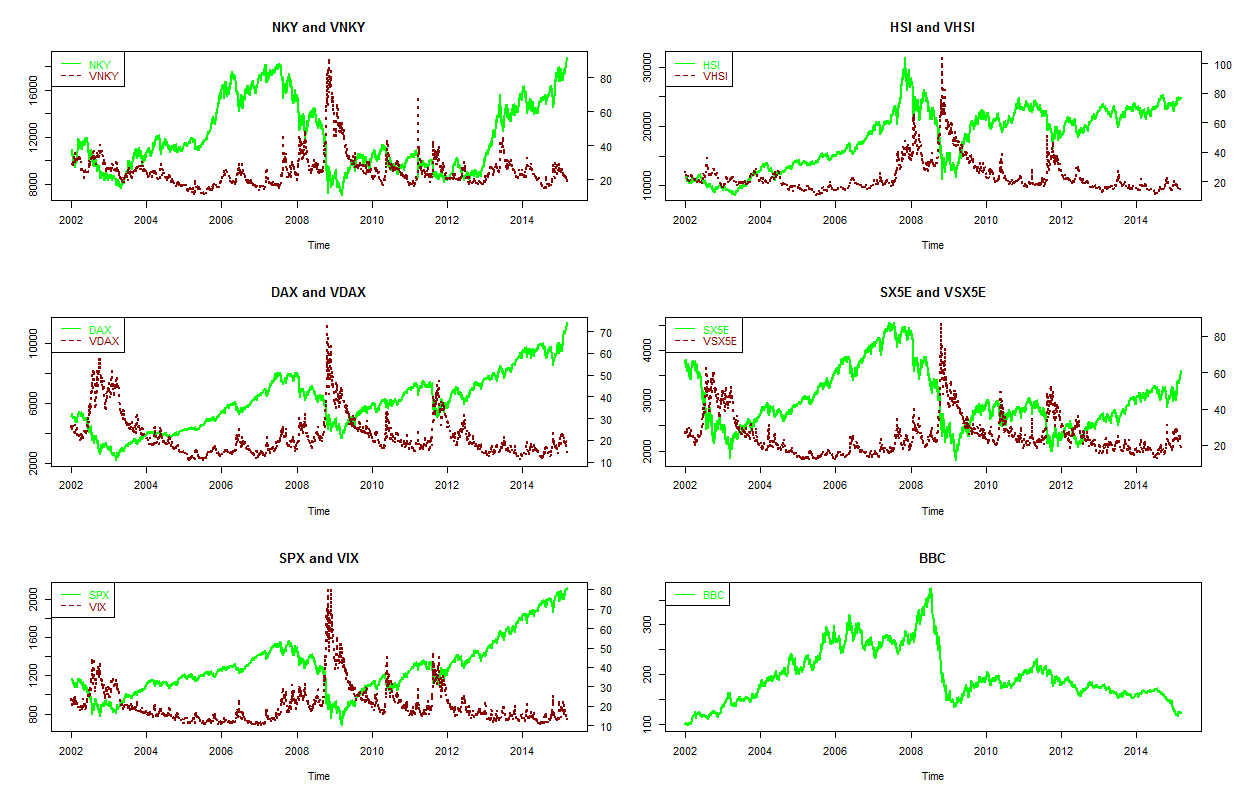}
\caption{Absolute levels of the considered time series.}
\label{levels}
\end{figure}

From a mathematical point of view, for such global analyses one faces the challenge of statistically describing higher-dimensional dependencies, which is usually tackled by either multivariate ARMA-GARCH setups or copula models. While the first ones might provide a fast and straightforward approach to analyze smaller data sets, parameters increase rather drastically for higher dimensions. Copulas on the other hand provide a neat way to circumvent such issues and therefore have found various applications in finance, e.g., in portfolio management (cf. \cite{Chollete11} for a recent application or \cite{patton2009} for a good overview). Several studies in the past invoked copula constructions to investigate either market dependencies, contagion or spillover effects as well, cf. \cite{Rodriguez07}, \cite{Okimoto08}, \cite{AD12} or \cite{BW15}. However, to the best of our knowledge, there is no full analysis about the common dependence structure of global equity and implied volatility indices. \cite{SN13} investigated a non-parametric tail dependence estimator but focussed only on bivariate pairs, while \cite{Kasch13} only considered global equity indices based on a variation of the DDC-MGARCH model. Most closely related to our work is probably \cite{peng2012} who applied a dynamic mixed copula approach with a focus on tail dependencies to analyze contagion effects between stock index movements and showed that financial shocks have a stronger effect on the relationship between volatility indices as on the one between equities. Additionally, they find that (asymmetric) tail dependencies tend to change and increase in post-crisis periods.

We shall build on their results but go a step further and directly incorporate potential regime switches in our analysis. The general idea of such setups based on Markov chains was introduced by \cite{Hamilton1989} and has recently been linked with copula models as well (cf. \cite{CholleteHV09}, \cite{daSilvaFilhoa12}, \cite{WWL13} or \cite{Stoeber2013}). Especially the approach based on regular vine copulas of \cite{Stoeber2013} is suited for our high-dimensional data set and we shall therefore build our analysis on their model setup. However, while they focussed on modeling price returns -- of FX rates to be precise -- we shall include volatility indices as well. In particular, we are interested in a Markov model with two states which we shall call the \enquote{normal} and \enquote{abnormal} regime. We are deliberately avoiding the usually applied wording of \enquote{non-crisis} and \enquote{crisis} as it is ex-ante not clear how our two states should look like. In fact, we will show that an asymmetric, high tail dependence relationship between equity and volatility is rather the \enquote{normal} state than a \enquote{crisis}.

\begin{table}[htbp]
\centering
  \begin{tabular}{ l | l | l }
  	\multicolumn{1}{l}{Shortcut}		& \multicolumn{1}{c}{ Index Description }									& Currency \\ \hline \hline
  	 \multicolumn{3}{c}{} \\
 \multicolumn{3}{c}{\textbf{Asia}} \\ \hline
    NKY 			& Nikkei-225 Stock Average Index									& JPY \\
        \hline
    VNKY 		& Implied Volatility Index of the NKY 											& JPY \\
            \hline
    HSI 			& Hang Seng Index 								& HKD \\
        \hline
    VHSI 		& Implied Volatility Index of the HSI  											& HKD \\
                   \hline
                     	 \multicolumn{3}{c}{} \\
   \multicolumn{3}{c}{\textbf{Europe}} \\ \hline
    DAX 		& Deutscher Aktien Index (German Stock Index) 	& EUR \\
        \hline
    VDAX 		& Implied Volatility Index of the DAX 											& EUR \\
            \hline
    SX5E 		& Euro Stoxx 50 Index  									       	& EUR \\
    \hline
    VSX5E 	& Implied Volatility Index of the SX5E								& EUR \\
                \hline
                	  	 \multicolumn{3}{c}{} \\
    \multicolumn{3}{c}{\textbf{USA}} \\ \hline
    SPX         & Standard and Poor's 500 	Index									& USD \\
        \hline
    VIX 			& Implied Volatility Index of the SPX 											& USD \\
                   \hline
    BBC 			& Bloomberg Commodity ex-Agriculture \& Livestock Index  & USD \\
        \hline
  \end{tabular}
   \caption{Considered indices separated by regions.}
   \label{data_description}
 \end{table}

To be precise, the present paper considers eleven indices from three major regions: Asia, Europe and the USA. In total, we have five equity indices as well as their corresponding implied volatility indices (cf. Table~\ref{data_description}). Additionally, one commodity index is taken into account: the Bloomberg Commodity ex-Agriculture and Livestock Index which is calculated mainly via futures trading in New York, Chicago and London. The considered time period covers roughly 13 years, starting in particular on 01.01.2002 and ending on 27.02.2015. Excluding non-trading days, this results in 3434 observations of daily closing prices in the respective domestic currency. The data source is Bloomberg and the absolute levels are depicted in Figure~\ref{levels}.

In a first step, we filter the marginal time series and transform them to uniformly distributed data on $[0,1]$ using ARMA-GARCH and ARMA-EGARCH models. We shall follow the procedure outlined in \cite{Beil} (who considered a similar but smaller data set) and choose the respective lags, GARCH-setup and error distribution in terms of log-likelihood, AIC, BIC, the $p$-value of the Ljung-Box test and a graphical evaluation of $qq$-plots. The selected setups can be found in Table~\ref{marginal_models}. The transformation of the obtained standardized residuals to uniformly distributed data is done by using the \textit{probability integral transform}: Let $Z$ be a random variable with distribution function $F_Z$, then $U:=F_Z(Z)$ is uniformly distributed on $[0,1]$.

\begin{table}[htbp]
\centering
  \begin{tabular}{ l | l | c }
  			& 														& Distribution \\
  		Index	& 		\multicolumn{1}{c|}{Marginal Model}			& of Innovations\\ \hline \hline
   NKY     & ARMA(3,2) - GARCH(1,1) &  GHYP \\
   VNKY     & ARMA(3,0) - GARCH(1,1) &  GHYP \\
   HSI      & ARMA(0,0) - GARCH(1,1) &  GHYP  \\
   VHSI      & ARMA(4,2) - GARCH(1,1) &  GHYP \\ \hline
       DAX 		& ARMA(0,0) - GARCH(1,1) &  GHYP \\
    VDAX 		& ARMA(0,0) - EGARCH(1,1) &  GHYP \\
   SX5E 		& ARMA(1,0) - GARCH(1,1) &  GHYP \\
    VSX5E 		& ARMA(3,0) - GARCH(1,1) &  GHYP \\ \hline
   SPX		& ARMA(3,0) - GARCH(1,1) &  GHYP \\
   VIX		& ARMA(3,0) - GARCH(1,1) &  GHYP \\
   BBC        & ARMA(0,0) - GARCH(1,1) &  GHYP
  \end{tabular}
   \caption{The selected models and distribution of innovations for the marginal time series where GHYP stands for the generalized hyperbolic distribution.}
   \label{marginal_models}
 \end{table}

The remainder of the paper is structured as follows. Section~\ref{sec2} will briefly review the construction of $R$-vine-copulas. Furthermore, the new concept of \emph{quarter tail dependence} (which is used to identify our two regimes later on) will be motivated and introduced. In Section~\ref{sec3} we provide a first, static application to our equity and volatility data, identifying plausible $R$-vine structures globally and for each continent individually. Section~\ref{sec4} summarizes the necessary preliminaries and results for regime-switching $R$-vine copulas from \cite{Stoeber2013}. Afterwards, in Section~~\ref{sec5}, we carry out an extensive rolling window analysis to determine suitable copula families for the \enquote{normal} and \enquote{abnormal} regimes. Different Markov-switching models are estimated, discussed and compared to their static counterparts. A brief summary closes the paper.

All calculations have been performed with the statistical software package R. For the marginal models fitted above, we applied the function \textit{ugarchfit} from the R-package \textit{rugarch} by \cite{rgarch}. All vine copulas computations, such as model selection, likelihood calculation and parameter estimation, are carried out via the R package \textit{VineCopula} by \cite{VineCopula}. The fitting of the Markov-switching $R$-vine copula models is based on the code of \cite{Stoeber2013}.
\quad\\

\section{Copula models without regime switches}\label{sec2}
In the following section we shall briefly review the main concepts of (regular vine) copulas, the core tool of our upcoming investigations, starting with the very general definition. Additionally, we introduce a new measure called \emph{quarter tail dependence} to capture tail asymmetries within our equity and volatility data set.
\quad\\

\subsection{Regular vine copulas}
For more details on (vine) copulas, we refer the interested reader to \cite{Joe1997}, \cite{Nelsen2006}, \cite{Kurowicka}, \cite{kj10}, \cite{csms12} or \cite{joe2014}. The general definition follows.

\begin{definition}
Let $d\geq 2$. A $d$-dimensional cumulative distribution function on $[0,1]^d$ with uniform, univariate marginal distributions is called a \emph{copula}. If the derivative of $C$ exists, its \emph{copula density} is given by
\begin{equation}
c(u_1,\ldots,u_d) = \frac{\partial^d}{\partial u_1 \cdots \partial u_d}C(u_1,\ldots,u_d)
\end{equation}
\end{definition}

The seminal theorem of \cite{Sklar1959} clarifies that for every multivariate distribution, a suitable copula exists.

\begin{theorem}[Sklar's theorem]
Let $d\geq 2$ and $X=(X_1, \dots, X_d)$ be a random vector with margins $F_1, \dots, F_d$ and a joint distribution function $F$. Then there exists a copula $C$, such that
\begin{equation} \label{Sklar}
F(x_1,\dots,x_d)=C(F_1(x_1), \dots, F_d(x_d)) \quad \forall x=(x_1, \dots, x_d)^T \in \mathbb{R}^d.
\end{equation}
If $F_1, \dots, F_d$ are continuous, then the copula $C$ is unique. Conversely, if $F_1, \dots, F_d$ are univariate distribution functions and $C$ is a $d$-dimensional copula, then $F$ defined via \eqref{Sklar} is a $d$-dimensional distribution function.
\end{theorem}

Straightforward examples which can be represented analytically even in higher dimensions are the classical Gauss or Student-$t$ copulas. However, for the purpose of modeling financial data, these might not always be a sensible choice: The Gaussian copula is not able capture tail dependence, while the Student-$t$ copula only allows for symmetric dependence in the lower and upper tails.

An elegant way of constructing more general multivariate distributions through conditioning using bivariate copulas was first introduced by \cite{BC01, BC02} based on earlier work of \cite{Joe1996} (see also \cite{csms12} for a recent summary). The resulting so-called \emph{pair copula construction (PCC)} which has been investigated in detail by \cite{Aas2009} and \cite{czado10} can be neatly illustrated in three dimensions by the following example: Let $f$ be the continuous density function of a 3-dimensional distribution function $F$ with marginals $F_1, F_2$ and $F_3$. Then we can decompose
\begin{align}\label{pcc}\nonumber
f\left(x_1, x_2, x_3\right) & = c_{13;2}\left(F_{1|2}\left(x_1|x_2\right),F_{3|2}\left(x_3|x_2\right)\right)\\
& \times c_{23}\left(F_2\left(x_2\right),F_3\left(x_3\right)\right)c_{12}\left(F_1\left(x_1\right),F_2\left(x_2\right)\right)\\\nonumber
& \times f_3\left(x_3\right)f_2\left(x_2\right)f_1\left(x_1\right),
\end{align}
where $c_{13;2}$ is the copula density associated with the distribution of $\left(X_1, X_3\right)$ given $X_2 = x_2$. In the following, we invoke the usually imposed \emph{simplifying assumption} which states that $c_{13;2}$ is independent of $x_2$. For a detailed discussion of its implication and potential restrictiveness we refer to \cite{hhaf10}, \cite{agn12a} and \cite{sjc13}. Also, the current non-simplified models are restricted to mostly three dimensions. In addition a dynamic formulation as we will propose allows to mitigate potential non-simplifying effects.

By repeated conditioning, the above outlined procedure can directly be generalized to higher dimensions. One has to keep in mind that decomposition~\eqref{pcc} is generally not unique. However, using graph-theoretical fundamentals a certain choice can be illustrated graphically in a compact and clear way (cf. \cite{BC01} and more generally \cite{Swamy}).

\begin{definition}
Let $d\geq 2$. A \emph{regular vine ($R$-vine)} is an ordered sequence of trees $\mathcal{V}=(T_1, \dots, T_{d-1})$ with ${T_i=(N_i, E_i)}$, $i\in\{1, \dots, d-1\}$, where $N_i$ is a set of nodes and $E_i$ a set of edges, such that
\begin{enumerate}
\item $N_1=\{1, \dots, d\}$, i.e. the first tree has nodes $1,\ldots,d$,
\item for $i\in\{2, \dots, d-1\}$ we have $N_i=E_{i-1}$, i.e. the nodes of $T_{i}$ are the edges of $T_{i-1}$,
\item if for $i\in\{1, \dots, d-2\}$ two nodes of $T_{i+1}$ are connected, the corresponding edges in $T_i$ have a common node \emph{(proximity condition)}.
\end{enumerate}
\end{definition}

If on each tree, every node has maximally two edges, we speak of a \emph{drawable vine ($D$-vine)} while if on each tree there exists one node, that is connected to all other nodes, we call this a \emph{canonical vine ($C$-vine)}.

\begin{figure}[htbp]
\centering
\includegraphics[height=3.05cm]{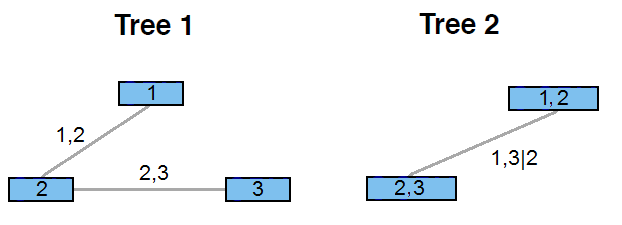}
\caption{Graphical representation of decomposition~\eqref{pcc}.}
\label{ex_trees}
\end{figure}

Coming back to our 3-dimensional example in \eqref{pcc}, the corresponding $R$-vine describing this structure is depicted by Figure~\ref{ex_trees}. In such cases, the corresponding copula $C$ is called a \emph{$R$-vine copula} and analogously a \emph{$D$-vine copula} or \emph{$C$-vine copula}. Since we considered only 3 dimensions, Figure~\ref{ex_trees} actually shows both, a $D$- and $C$-vine structure.

Having outlined how higher-dimensional copulas can be constructed using bivariate distributions, we still have to choose suitable two-dimensional copulas. In general, there are two major bivariate copula families: Elliptical and Archimedean. The first group contains the bivariate Gauss and Student-$t$ copulas while examples for the Archimedean ones are the Clayton, Gumbel, Frank or Joe copulas. A very neat overview is presented in \cite{BC15}. In the following, we will consider Gauss and Student-$t$ copulas which are the most common bivariate models used in financial applications. Due to their property of allowing asymmetric tail dependence, we shall further focus on the Gumbel copula and its counter-clockwise rotations defined for $\theta\in[1,\infty)$ and $u,v,\in[0,1]$ via
\begin{eqnarray}\nonumber
C_{{\mathrm{Gumbel}}}(u,v) &=& \exp\left\{-\left[\left(-\log(u)\right)^\theta + \left(-\log(v)\right)^\theta\right]^{\frac{1}{\theta}}\right\}\\\nonumber
C_{\mathrm{Gumbel90}}(u,v)&=&v - C_{{\mathrm{Gumbel}}}(1-u,v)\\
C_{\mathrm{Gumbel180}}(u,v)&=&u+v-1+C_{{\mathrm{Gumbel}}}(1-u,1-v)\\\nonumber
C_{\mathrm{Gumbel270}}(u,v)&=&u-C_{{\mathrm{Gumbel}}}(u,1-v)
\end{eqnarray}

We finish this subsection by recalling Kendall's $\tau$, the most widely used dependence measure in copula theory due to its property of being invariant with respect to strictly monotone transformations of the marginal distributions. Therefore the ARMA-GARCH filtering in Section~\ref{sec1} does not influence Kendall's $\tau$. In fact, if $C$ is the corresponding copula for a bivariate random vector $(X,Y)^T$, then we have
\begin{equation}\label{Taus}
\tau(X,Y)=4\left[\int_{[0,1]^2} C(u_1,u_2) d C(u_1, u_2)\right] - 1.
\end{equation}
%

For given sample data $(x_1,y_1)^T,\ldots(x_n,y_n)^T$, $n\geq 2$, an estimate of Kendall's $\tau$ is given by
\begin{equation}
\hat{\tau}(X,Y) = \frac{N_c-N_d}{\sqrt{N_c+N_d+N_X}\cdot\sqrt{N_c+N_d+N_Y}}.
\end{equation}
where $N_d$ and $N_c$ are the numbers of discordant and concordant pairs, respectively, while $N_X$ ($N_Y$) denotes the amount of the pairs with $x_i = x_j$ ($y_i = y_j$) where $i,j\in\{1,\ldots,n\}$, $i\neq j$ (\cite{Kurowicka}). For a data set without ties, there are simpler formulas, as well. We shall call $\hat{\tau}(X,Y)$ the empirical Kendall's $\tau$.

\quad\\

\subsection{Quarter tail dependence}
As pointed out before, the main aim of this paper is to identify \enquote{normal} and \enquote{abnormal} dependence regimes within the relationship between equity and implied volatility indices. One can expect that in times of market turmoil, especially the dependence structure in the left tails between stocks increases. Similarly, the same holds true for the right tails of two volatility indices. Both effects can be captured by the well-known (bivariate) lower and upper tail dependence coefficients defined as follows:

\begin{definition}[Tail dependence coefficients]\label{tdc}
For a copula $C$ of a random vector $(U,V)^T$ with marginal distribution functions $F_1$ and $F_2$, we define its \emph{upper} and \emph{lower tail dependence coefficients (TDCs)} via
\begin{align}
	\lambda_{U}(C)&= \lim_{u \nearrow 1}  \mathbb{P} (V > F_2^{-1}(u)| U > F_1^{-1}(u)) = \lim_{u \nearrow 1} \frac{1-2u+C(u,u)}{1-u},\\
	\lambda_{L}(C) &= \lim_{u \searrow 0}  \mathbb{P} (V \leq F_2^{-1}(u)| U \leq F_1^{-1}(u)) = \lim_{u \searrow 0} \frac{C(u,u)}{u}.
	\end{align}
\end{definition}
The measures $\lambda_{U}(C)$ and $\lambda_{L}(C)$ are able to capture (bivariate) relationships within the first and third quadrant of $[0,1]^2$. Thus, not every copula family has a non-zero upper or lower tail dependence, cf. Table~\ref{tdc2}. Consider, e.g., the Gumbel copula and its rotations: all four copulas describe different tail dependence properties. The upper tail dependence coefficient is non-zero only for the Gumbel copula while the lower tail dependence coefficient vanishes for all but Gumbel180.

\begin{table}[!h]
\centering
\begin{tabular}{l l l }
Copula family & $\lambda_L$ & $\lambda_U$  \\ \hline
Gauss & - & - \\
Student-$t$ &  $2t_{v+1}\left(-\sqrt{v+1}\sqrt{\frac{1-\theta}{1+\theta}}\right)$ & $2t_{v+1}\left(-\sqrt{v+1}\sqrt{\frac{1-\theta}{1+\theta}}\right)$\\
Gumbel & - & $2 - 2^{1/{\theta}} $\\
Gumbel90 & - & - \\
Gumbel180 &  $2 - 2^{1/{\theta}}$ & - \\
 Gumbel270 & - & - \\
\end{tabular}
\caption{Tail dependence coefficients for different copula families. $\theta$ is the copula parameter for the Gumbel/Student-$t$ families and $v$ denotes the degrees of freedom of the Student-$t$ copula.}
\label{tdc2}
\end{table}

However, when it comes to the (asymmetric) relationship between equity and volatility, we would expect Gumbel90 or Gumbel270 to be fitting better as the \enquote{normal} relation observed in financial markets is often the following: On the one side, turmoil in equity markets is usually accompanied by a steep rise in volatility levels. On the other side, booming markets tend to be associated with declining volatility even though this effect is usually not that strong. As shown above, this asymmetric relationship can not be captured by the lower or upper tail dependence coefficients.

Additionally, we are considering a commodity index, as well. In a time period of stagnating or declining global growth, which would be accompanied by low or negative equity and high volatility returns, one can expect that the demand for, e.g., industrial metals and energy tends to fall heavily. As consequence, the tail dependence for a commodity-volatility index pair should be analogues to an equity-volatility pair.

There are various ways to describe this kind of asymmetric dependence structure (cf. \cite{SN13} who considered tail copulas). For our purposes, we shall slightly tweak the definition of the tail dependence coefficients to get suitable measures for the above described relationships:

\begin{table}[htbp]
\centering
\begin{tabular}{| l l || l l | }
\hline
\multicolumn{2}{|c||}{$\lambda_{QTD}(U, V, C):=\lambda_{QTD}^2(C)$} & \multicolumn{2}{c|}{$\lambda_{QTD}(U, V, C):=\lambda_{QTD}^1(C)$} \\
\multicolumn{2}{|c||}{for $(U,V)^T\in\{(Eq, Vol)^T, (Cmd, Vol)^T\}$:}  & \multicolumn{2}{c|}{for $(U,V)^T=(Vol,Vol)^T$:} \\
\multicolumn{2}{|c||}{}& \multicolumn{2}{c|}{}\\
\multicolumn{2}{|c||}{\footnotesize{$\lambda_{QTD}^2(C)=\left\{\begin{array}{cl} \lambda_U(St(-\theta,\nu)), & if ~ C=St(\theta, \nu) \\ & \mathrm{with} ~ \tau<0 \\ \lambda_U(Gu(-\theta)), & if ~ C=Gu90(\theta)\\ 0, & \mbox{else} \end{array}\right. $}}& \multicolumn{2}{c|}{\footnotesize{$ \lambda_{QTD}^1(C)=\lambda_U(C(\theta,\nu))$}}\\
\multicolumn{2}{|c||}{}& \multicolumn{2}{c|}{}\\
\multicolumn{2}{|c||}{ \includegraphics[height=4cm]{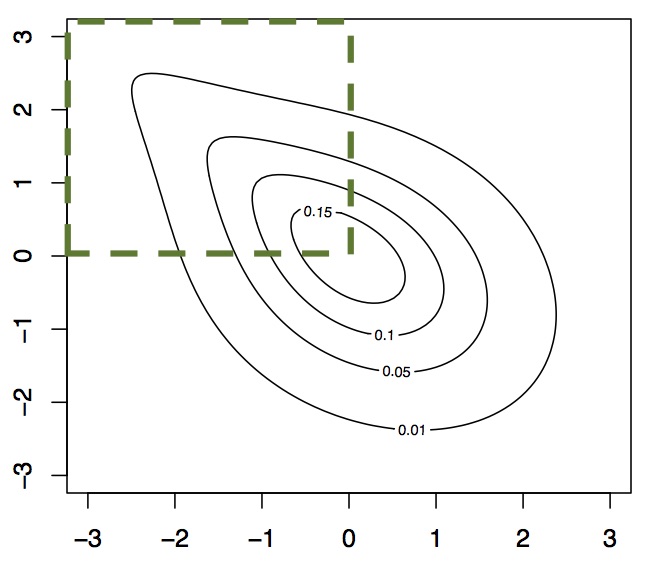}} & \multicolumn{2}{c|}{\includegraphics[height=4cm]{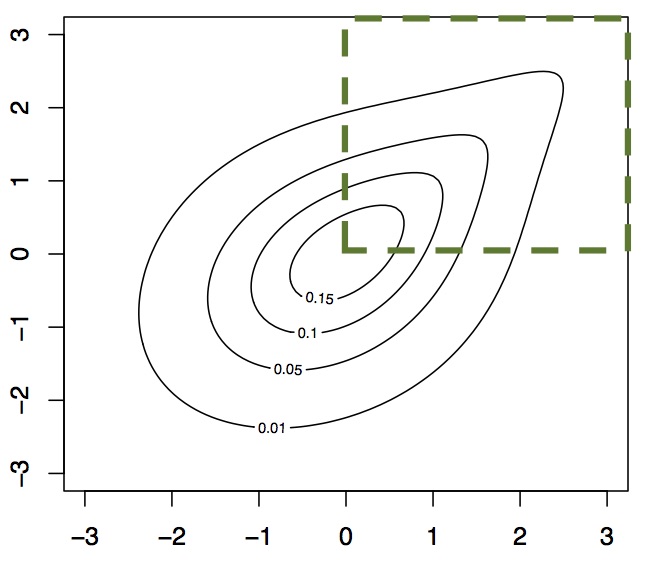}} \\
\multicolumn{2}{|c||}{\textit{"Eq/Cmd goes down, Vol up"}}  & \multicolumn{2}{c|}{\textit{"Both Vol indices go up"}} \\
\hline \hline
\multicolumn{2}{|c||}{$\lambda_{QTD}(U, V, C):=\lambda_{QTD}^3(C)$} & \multicolumn{2}{c|}{$\lambda_{QTD}(U, V, C):=\lambda_{QTD}^4(C)$}\\
\multicolumn{2}{|c||}{for $U,V\in\{Eq,Cmd\}$:}  & \multicolumn{2}{c|}{for $(U,V)^T\in\{(Vol, Eq)^T, (Vol, Cmd)^T\}$:} \\
\multicolumn{2}{|c||}{}& \multicolumn{2}{c|}{}\\
\multicolumn{2}{|c||}{\footnotesize{$ \lambda_{QTD}^3(C)=\lambda_L(C(\theta,\nu))$}}& \multicolumn{2}{c|}{\footnotesize{$\lambda_{QTD}^4(C)=\left\{\begin{array}{cl} \lambda_U(St(-\theta,\nu)), & if ~ C=St(\theta, \nu) \\ & \mathrm{with} ~ \tau<0 \\ \lambda_U(Gu(-\theta)), & if ~ C=Gu270(\theta)\\ 0, & \mbox{else} \end{array}\right. $}}\\
\multicolumn{2}{|c||}{}& \multicolumn{2}{c|}{}\\
\multicolumn{2}{|c||}{\includegraphics[height=4cm]{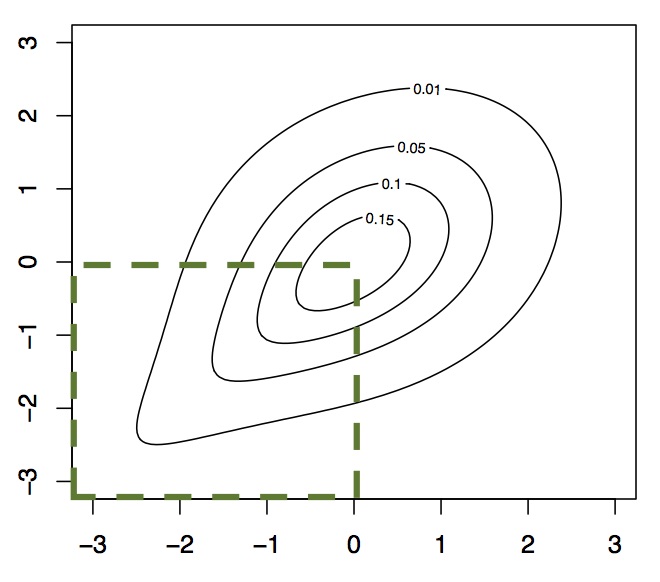}} & \multicolumn{2}{c|}{\includegraphics[height=4cm]{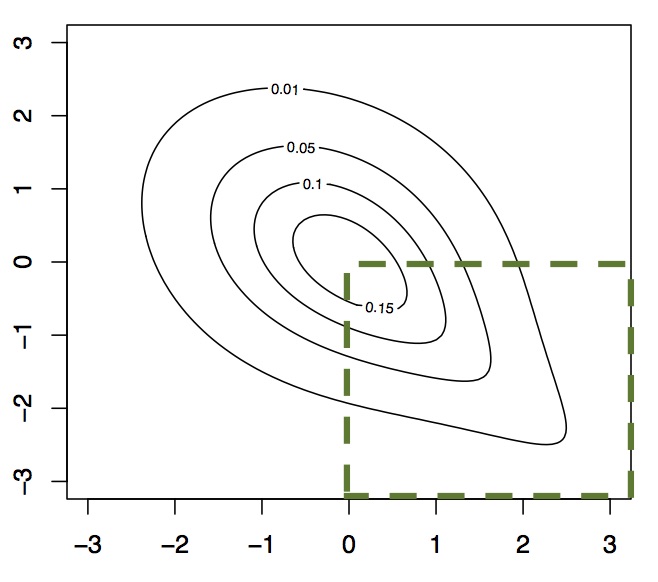}} \\
\multicolumn{2}{|c||}{\textit{"Both Eq and Cmd/Eq go down"}}  & \multicolumn{2}{c|}{\textit{"Eq/Cmd goes down, Vol up"}} \\ \hline
\end{tabular}
\caption{Quarter tail dependence $\lambda_{QTD}(X, Y, C)$ for different index pairs and copula families: Gauss, Student-$t$ and Gumbel including rotations. The graphical example is based on contour plots of the Gumbel copula and its rotations with Kendall's $\tau$ of 0.3 and -0.3 respectively.}
\label{qtd}
\end{table}

\begin{definition}[Quarter tail dependence]\label{qtddef}
Let $(U,V)^T$ be a random vector with uniformly distributed marginals. We specifically assume, that $(U,V)^T$ are correspondingly transformed returns of equity (Eq), volatility (Vol) or commodity (Com) indices, using from now on the notation $U,\;V \in (Eq, Vol, Cmd)$. Furthermore let the copula $C$ of $(U,V)^T$ be either Gauss (Ga), Student-t (St-t), Gumbel or one of its rotations (Gu, Gu90, Gu180 or Gu270). Then the \emph{quarter tail dependence (QTD)} is defined by
\begin{equation}
\lambda_{QTD}(U,V,C)=\left\{\begin{array}{cl} \lambda_{QTD}^1 (C), & if ~ (U,V)^T=(Vol,Vol)^T \\ \lambda_{QTD}^2 (C), & if ~ (U,V)^T\in\{(Eq, Vol)^T, (Cmd, Vol)^T\} \\  \lambda_{QTD}^3 (C), & if ~ (U,V)^T\in\{(Eq, Eq)^T, (Cmd, Eq)^T, (Eq, Cmd)^T\} \\ \lambda_{QTD}^4 (C), & if ~ (U,V)^T\in\{(Vol, Eq)^T, (Vol, Cmd)^T\}  \end{array}\right.
\end{equation}
where
\begin{eqnarray}\nonumber
\lambda_{QTD}^1(C)&=&\lambda_U(C), \\\nonumber
\lambda_{QTD}^2(C)&=&\begin{cases}\lambda_U(St(-\theta,\nu)), & if ~ C=St(\theta, \nu) ~ with ~ \tau<0 \\ \lambda_U(Gu(-\theta)), & if ~ C=Gu90(\theta)\\ 0, & \mbox{else}, \end{cases} \\
\lambda_{QTD}^3(C)&=&\lambda_L(C) , \\\nonumber
\lambda_{QTD}^4(C)&=&\begin{cases}\lambda_U(St(-\theta,\nu)), & if ~ C=St(\theta, \nu) ~ with ~ \tau<0 \\ \lambda_U(Gu(-\theta)), & if ~ C=Gu270(\theta)\\ 0, & \mbox{else}, \end{cases}
\end{eqnarray}
where $\lambda_L$ and $\lambda_U$ denote the lower and upper tail dependence coefficients as stated in Definition~\ref{tdc} and given in Table~\ref{tdc2} for the chosen copula families. Furthermore $\theta$ is the Gumbel/Student-$t$ copula parameter and $\nu$ denotes number of degrees of freedom in case of the Student-$t$ copula.
\end{definition}

As one can see, the (classical) upper tail dependence coefficient, which describes the dependence in the first quadrant, is used for the calculation of the QTD for a volatility-volatility pair. For an equity-equity as well as an equity-commodity pair we are interested in the third quadrant, which is measured using the lower tail dependence coefficient. Finally, an equity-volatility or a commodity-volatility relationship is located in the second and fourth quadrant and captured by $\lambda_{QTD}^2(C)$ and $\lambda_{QTD}^4(C)$. An illustration of the QTD-definition is given by Table~\ref{qtd}.
\quad\\

\section{Identifying plausible $R$-vine tree structures}\label{sec3}
\begin{figure}[htbp]
\centering
\includegraphics[height=7.5cm]{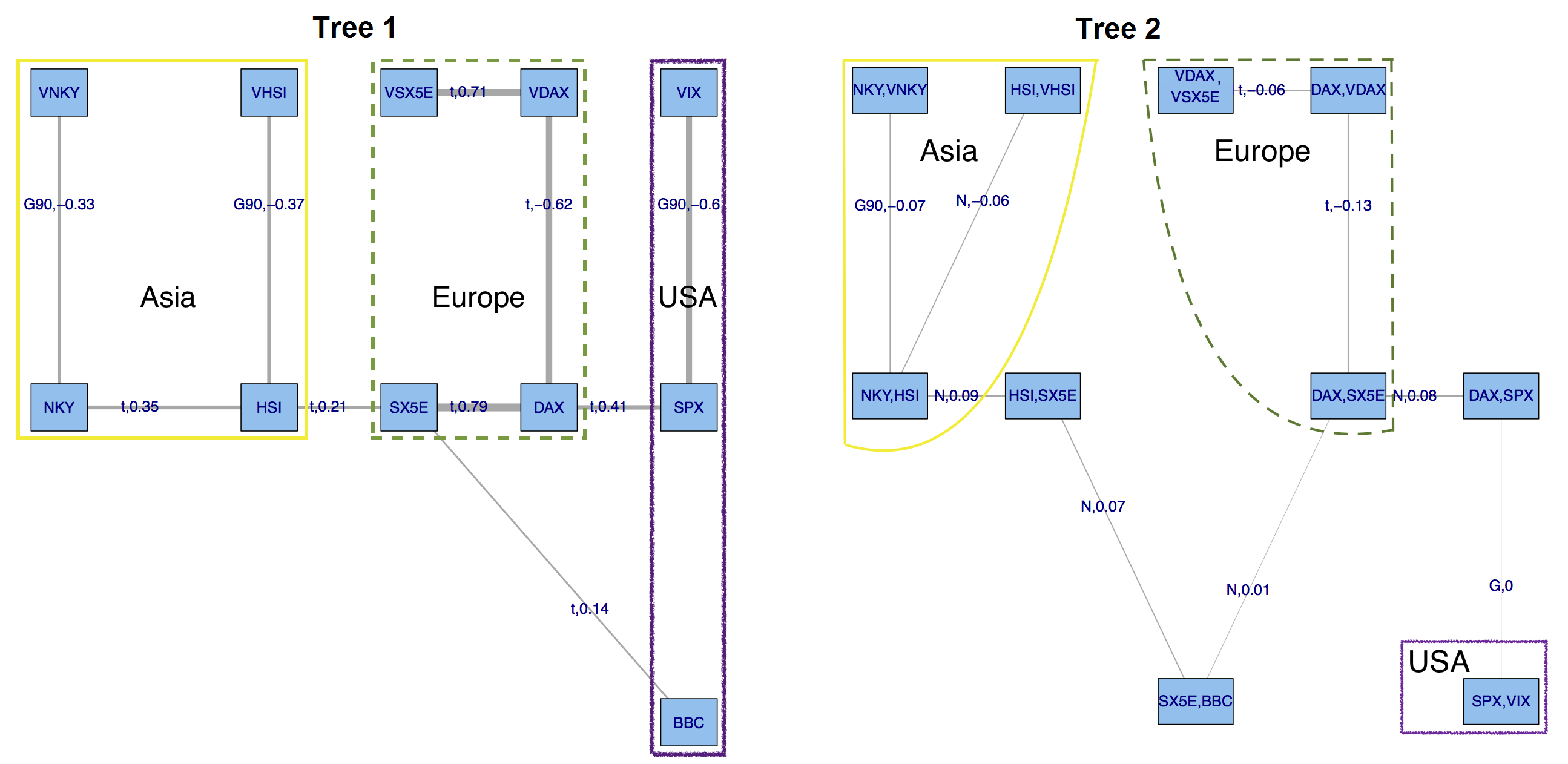}
\caption{Model (1-dependent): first and second tree, corresponding bivariate copula families and values of Kendall's $\tau$.}
\label{trees_main}
\end{figure}

In this section we shall provide a first, static look, at our index data. Considering the filtered return series from Section~\ref{sec1}, we firstly apply the $R$-vine structure selection technique of \cite{Dissmann2013} (which is basically connecting those nodes with the highest dependence measured by Kendall's $\tau$), i.e. no restrictions on tree structure and chosen copula families (Gaussian (Ga), Student-$t$ (St-$t$) and Gumbel (Gu) including all rotations (Gu90, Gu180 and Gu270)) are imposed. We shall call the resulting $R$-vine model (1-dependent) and its first two trees are depicted in Figure~\ref{trees_main}. The upcoming static global models with pre-defined tree structure will be denoted by \enquote{(2-...)} while we use \enquote{(3-...)} for the final regime switching models. The wording \enquote{dependent} or \enquote{independent} shall indicate whether we allow for a potential dependence structure between continents or not.

As we can see, the equity indices connect directly with their corresponding volatility counterparts indicating a strong dependence structure between both. On a more macro scale, this also holds true for the considered geographical regions as indices from the same continent seem to \enquote{stick together} - the commodity index being the only exception which could be explained by the fact that BBC is a global index, with underlying futures prices determined in New York, Chicago and London. Additionally and as expected, Kendall's $\tau$ indicates a positive dependence between equity-equity and equity-commodity and a negative one for equity-volatility.

Apart from BBC being linked with an European index, the continents are mainly connected by two pairs of equity indices, namely HSI-SX5E and DAX-SPX, which can be neatly explained by considering global trading hours (cf.  Table~\ref{trading_hours}).

\begin{table}[h]
\centering
\begin{tabular}{ l | c }
Stock Exchange & Opening - Closing Time (UTC) \\ \hline
\multicolumn{2}{c}{\textbf{Asia}} \\ \hline
Tokyo Stock Exchange & 00:00 - 06:00 \\
Hong Kong Stock Exchange & 01:30 - 08:00 \\\hline
\multicolumn{2}{c}{\textbf{Europe}} \\ \hline
Frankfurt Stock Exchange (Xetra) & 08:00 - 16:30 \\
Eurex Exchange (for SX5E index options) & 08:00 - 16:30 \\\hline
\multicolumn{2}{c}{\textbf{USA}} \\ \hline
New York Stock Exchange & 14:30 - 21:00 \\\hline
\end{tabular}
\caption{International trading hours (ignoring potential lunchbreaks).}
\label{trading_hours}
\end{table}

Finally, the level of connection measured by Kendall's $\tau$ is seen to be lower on the second than on the first tree while the dependence structures within each region seem to be stronger than those between the continents. For the later analysis, it suggests a more detailed investigation of the behavior within and between continents.

\begin{figure}[htbp]
\centering
\includegraphics[height=12cm]{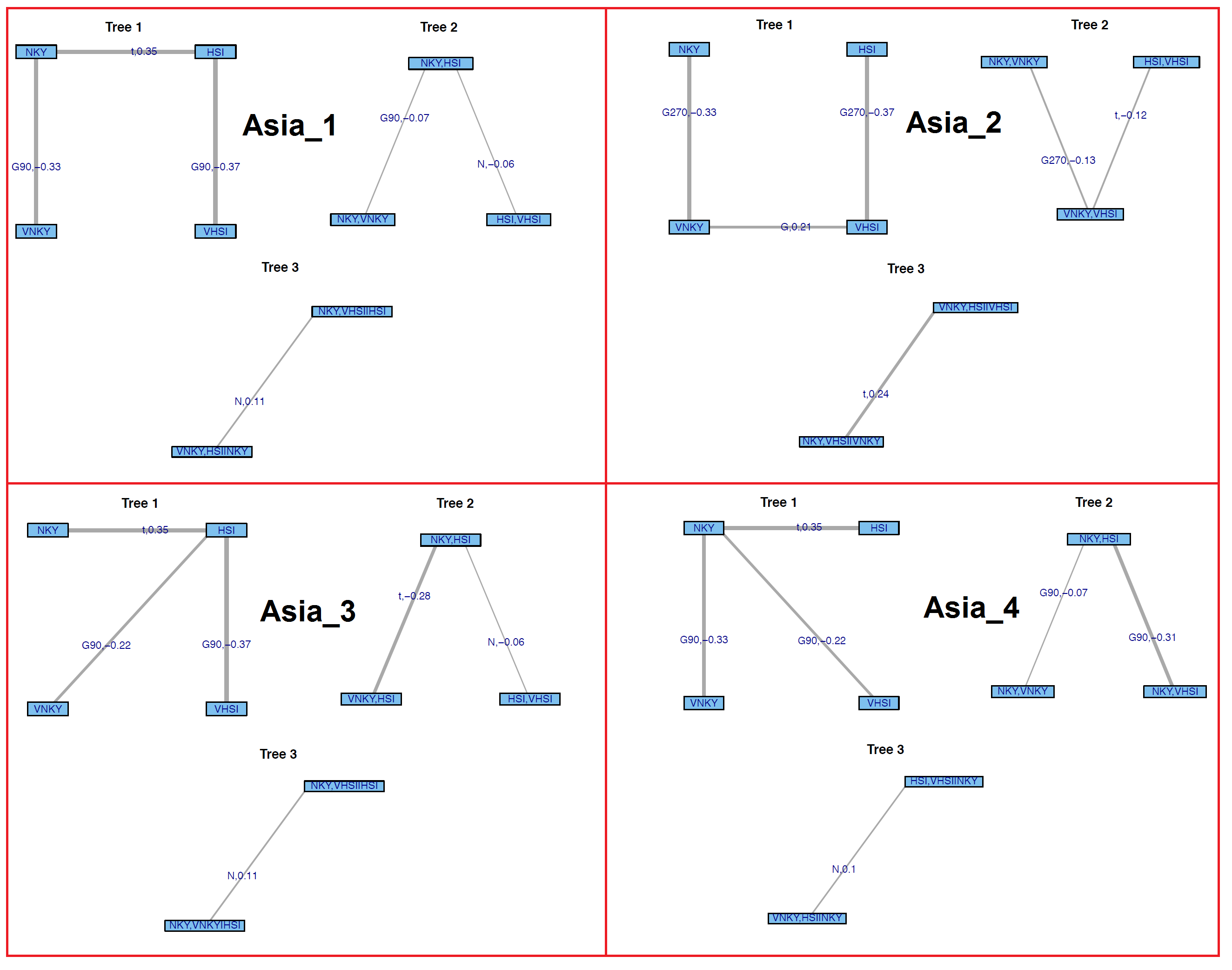}
\caption{Tree structures for models Asia$_1$-Asia$_4$ and values of Kendall's $\tau$.}
\label{Asia_trees}
\end{figure}

Therefore, to prepare a potential (and plausible!) pre-defined tree structure for our regime switching setup, we start by investigating all three intra-regional dependencies separately. Only in a second step, we shall add a global tree structure. Starting with Asia, we consider four setups, namely Asia$_1$-Asia$_4$ depicted in Figure~\ref{Asia_trees}. The first two models, Asia$_1$ and Asia$_2$, are $D$-vines, in which the connections are either made through equities (NKY-HSI) or volatilities (VNKY-VSHI), respectively. In the $C$-vine model Asia$_3$ the central node is the equity index HSI at the Hong Kong stock exchange while in Asia$_4$ this role is taken over by the Japanese NKY. From now on, for a given tree structure, estimation and copula selection is carried out based on maximum likelihood and AIC using the function \emph{RVineCopSelect} out of the \emph{VineCopula} package by \cite{VineCopula}.

\begin{figure}[htbp]
\centering
\includegraphics[height=12cm]{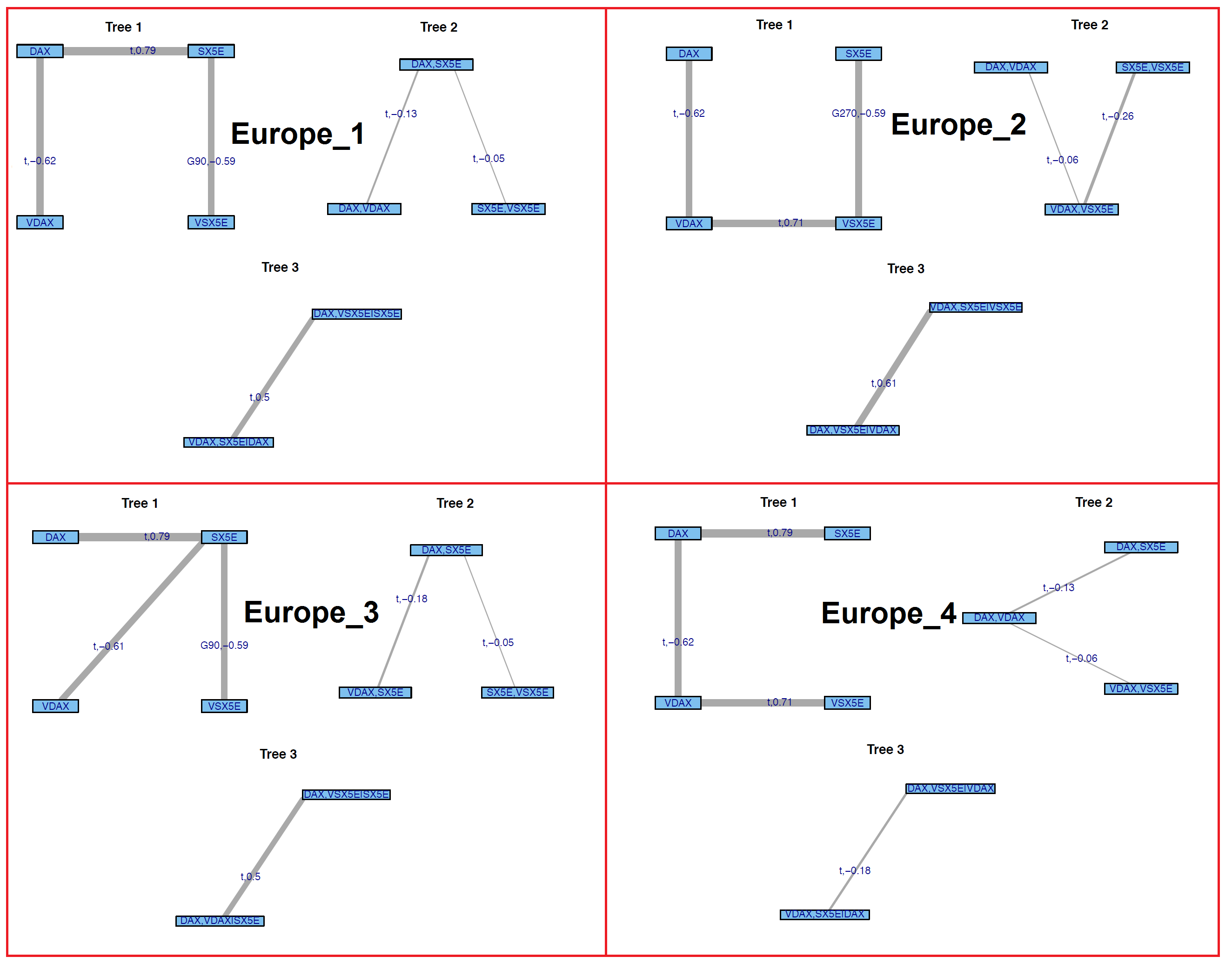}
\caption{Tree structures for models Europe$_1$-Europe$_4$ and values of Kendall's $\tau$.}
\label{EU_trees}
\end{figure}

Europe$_1$ and Europe$_2$ are build analogously to the ones in the Asian region. The central node of the $C$-vine Europe$_3$ is chosen to be the broader European equity index SX5E. As the DAX is partially included in the SX5E, Europe$_4$ is taken to be another $D$-vine for which the \enquote{middle} nodes are the DAX and its implied volatility index VDAX. Another reason for this particular setup is that this structure was also chosen in (1-dependent) and we want to test whether this is optimal when Europe is analyzed separately. All models are depicted in Figure~\ref{EU_trees}.

\begin{figure}[htbp]
\centering
\includegraphics[height=3.05cm]{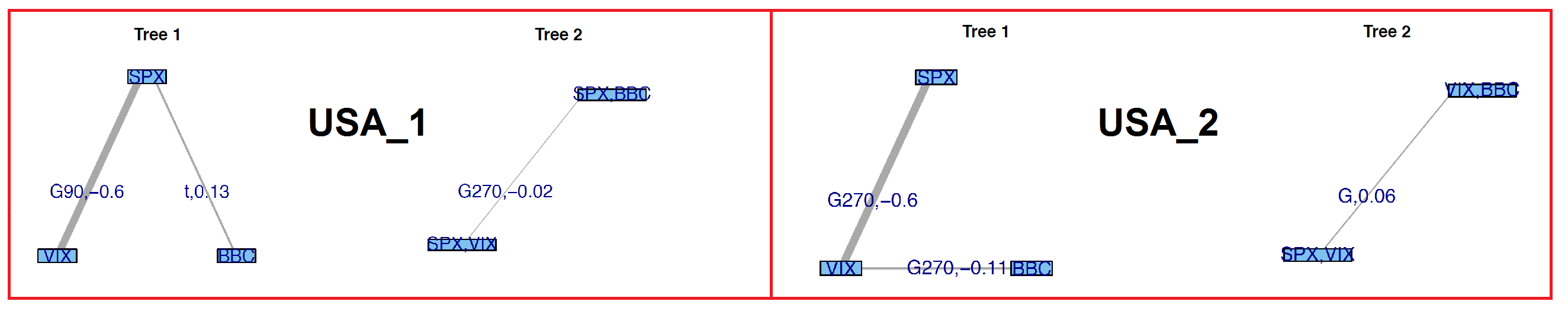}
\caption{Tree structures for models USA$_1$-USA$_2$ and values of Kendall's $\tau$.}
\label{USA_trees}
\end{figure}

Finally, the considered structures for North-America, USA$_1$ and USA$_2$, are shown in Figure~\ref{USA_trees}. Both are $C$-vine models where the central node in the first tree is either the SPX or the VIX.

For all setups, the log-likelihood and information criteria are presented in Table~\ref{model_comparison}. Generally, the $D$-vine models Asia$_1$, Europe$_1$ and USA$_1$ deliver the highest log-likelihoods and the best AIC and BIC values as well.

\begin{figure}[htbp]
\centering
\includegraphics[height=7.5cm]{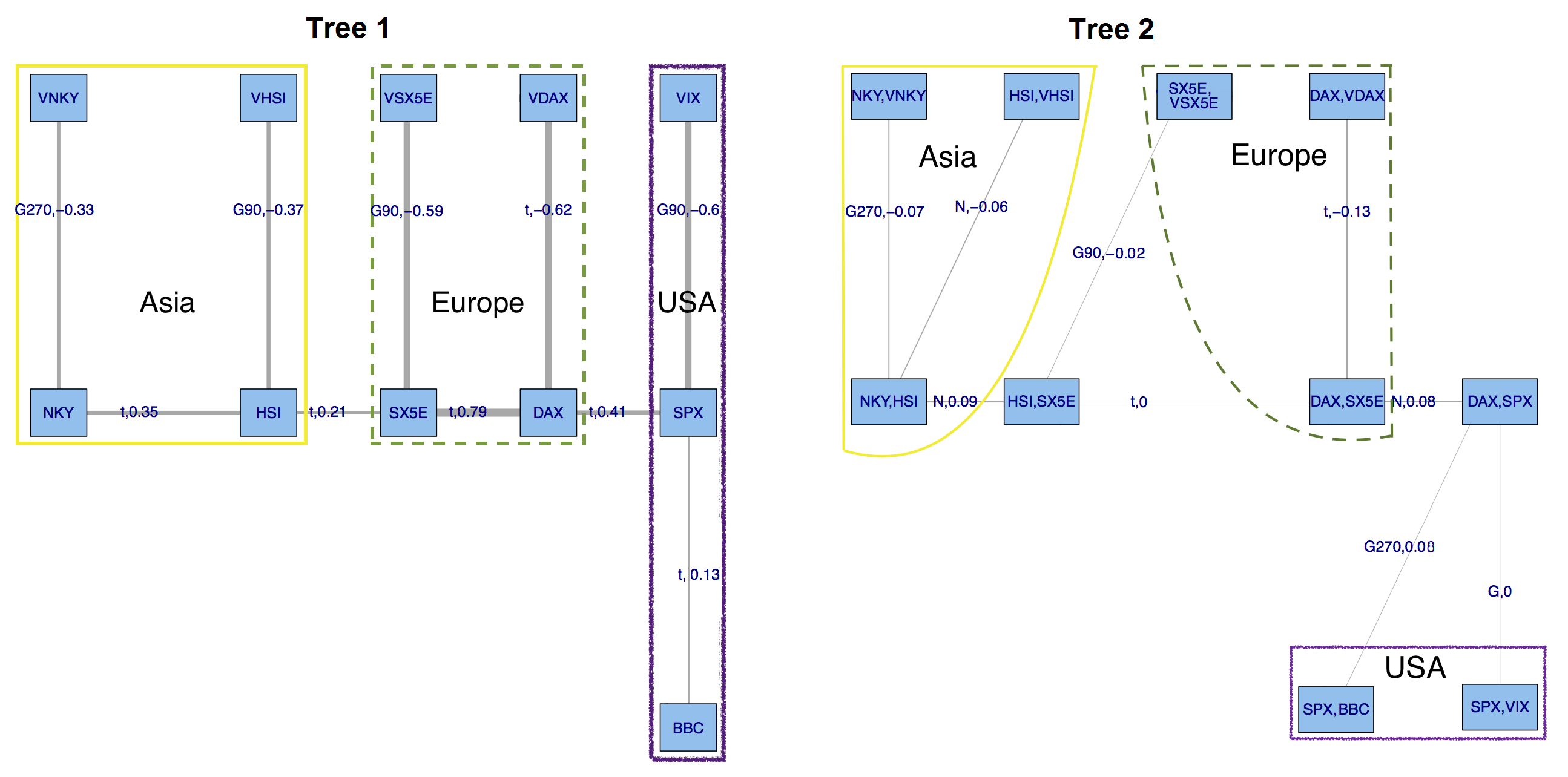}
\caption{Model (2-dependent): first and second tree, corresponding bivariate copula families and values of Kendall's $\tau$.}
\label{model2d}
\end{figure}

\begin{table}[htbp]
\centering
\begin{tabular}{ l | r | r | r }
& Log-likelihood &\multicolumn{1}{ c| }{AIC}  &\multicolumn{1}{c }{BIC} \\ \hline \hline
Model Asia$_1$ 		&\textbf{2,027.61}  &\textbf{-4,041.22} &\textbf{-3,998.24}\\
Model Asia$_2$		&  2,014.13 &-4,012.25& -3,963.12\\
Model Asia$_3$ 		& 1,997.10 &-3,978.21& -3,929.08\\
Model Asia$_4$ 		& 1,973.17 &-3,932.35& -3,889.36\\ \hline
Model Europe$_1$ 		&\textbf{8,973.48}  &\textbf{-17,924.95} &\textbf{-17,857.40}\\
Model Europe$_2$		&  8,758.88& -17,495.76& -17,428.21\\
Model Europe$_3$ 		& 8,964.05& -17,906.11& -17,838.55\\
Model Europe$_4$ 		& 8,944.14& -17,864.27 &-17,790.58\\ \hline
Model USA$_1$ 		& \textbf{2,017.63}&  \textbf{-4,027.26} & \textbf{-4,002.69}\\
Model USA$_2$ 		& 2,009.69&  -4,013.37 & -3,994.95\\ \hline
\end{tabular}
\caption{Log-likelihood, AIC and BIC for the considered $R$-vine models Asia$_1$-Asia$_4$, Europe$_1$-Europe$_4$ and USA$_1$-USA$_2$ for the geographical regions. Combining the bold models leads to (2-independent).}
\label{model_comparison}
\end{table}

Finally, we combine the previous results to set up a global model: Within each region, the structure of the first trees is inherited from Table~\ref{model_comparison} while the connections between continents are made via the same pairs as in (1-dependent), i.e. HSI-SX5E and DAX-SPX, resulting in model (2-dependent) depicted by Figure~\ref{model2d}. In particular, in terms of log-likelihood, as well as AIC and BIC, the new tree structure seems to be slightly better than (1-dependent) as summarized by Table~\ref{model_comparison_all}. Not very surprisingly, it also outperforms (2-independent) which is build from Asia$_1$, Europe$_1$ and USA$_1$ by assuming independent continents.

\begin{table}[htbp]
\centering
\begin{tabular}{ l | r | r | r }
& Log-likelihood &\multicolumn{1}{ c| }{AIC}  &\multicolumn{1}{c }{BIC} \\ \hline \hline
(1-dependent) 		&14,578.12 &-29,010.24 &-28,561.94\\
(2-independent) 		&  13,018.72&  -25,993.44 & -25,858.33\\
(2-dependent) 		& 14,618.70& -29,091.40& -28,643.09\\ \hline
\end{tabular}
\caption{Log-likelihood, AIC and BIC for the considered $R$-vine models (1-dependent), (2-independent) and (2-dependent).}
\label{model_comparison_all}
\end{table}
\quad\\

\section{Markov-switching $R$-vine copula models}\label{sec4}

In this section we review the regime switching setup based on $R$-vine copulas from \cite{Stoeber2013}, which will later on be used to model potential changes within the dependence structure of our index data set. As we are in particular interested in describing a \enquote{normal} and an \enquote{abnormal} market state, we will for ease of notation already now restrict ourselves to two different regimes.
The general idea of Markov-switching models, initially introduced by \cite{Hamilton1989}, is based on a latent Markov chain that governs the dependence structure of a time series. The later one shall be given in our case by $\{\textbf{U}_t=(U^1_{t},\dots, U^d_{t})\}_{t\in\{1, \dots, T\}}$ with $\{U^i_{t}\}_{t\in\{1, \dots, T\}}$ being i.i.d. uniform.

To be precise, let $\{S_t\}_{t=1, \dots, T}$ be a homogeneous, discrete-time Markov chain, which can take two different states $k=1$ and $k=2$, which we also call \emph{regimes}. Later on, $k=1$ will represent the \enquote{normal} state while $k=2$ will describe the \enquote{abnormal} one. The transition matrix shall be denoted by $\bm{\mathrm{P}}$, having elements given for $k,k'\in\{1,2\}$ by $P_{k,k'}:=P(S_t=k'|S_{t-1}=k)$, i.e. the transition probabilities from state $k$ at time $t-1$ to state $k'$ at time $t$. Finally, consider two different $R$-vine copula models $(\mathcal{V}^1, C^1, \bm{\theta}^1)$ and $(\mathcal{V}^2, C^2, \bm{\theta}^2)$ where $\mathcal{V}^i$, $i=\{1,2\}$, are the respective tree structures and $C^i$ the corresponding multivariate copulas having parameters stored in the vector $\bm{\theta}^i$.

Our two-regime \emph{Markov Switching $R$-vine (MS-RV) model} is therefore given by two $R$-vine copula specifications and the $2\times 2$ transition matrix $\bm{\mathrm{P}}$ describing the underlying Markov chain. The conditional copula density is given by
\begin{equation}\label{conddens}
c(\bm{\mathrm{u}}_t|(\mathcal{V}^i, C^i, \bm{\theta}^i)_{i\in\{1,2\}}, S_t)=\sum_{k=1}^2 1_{\{k\}}(S_t) \cdot c(\bm{\mathrm{u}}_t|(\mathcal{V}^k, C^k, \bm{\theta}^k)),\quad \bm{\mathrm{u}}_t\in[0,1]^d.
\end{equation}
Following \cite{Stoeber2013}, the full likelihood function of the MS-RV copula model can be decomposed into conditional densities via
\begin{eqnarray}\nonumber
f(\bm{\mathrm{u}}_1,\ldots \bm{\mathrm{u}}_T|\bm{\theta}^1, \bm{\theta}^2, \bm{\mathrm{P}})&=&\left[\sum_{k=1}^2 f(\bm{\mathrm{u}}_1|S_1=k, \bm{\theta}^k) P (S_1=k |\bm{\mathrm{P}}) \right]\\ &&\times \displaystyle\prod_{t=2}^{T} \left[\sum_{k=1}^2 f(\bm{\mathrm{u}}_t|S_t=k, \bm{\theta}^k) P(S_t=k |\bm{\mathrm{u}}_{1:(t-1)}, \bm{\mathrm{P}}) \right]
\end{eqnarray}
where $\bm{\mathrm{u}}_{1:t}:=(\bm{\mathrm{u}}_1,\ldots, \bm{\mathrm{u}}_t)$ for $t=\{1,\ldots,T\}$. To tackle the issue of maximizing the above likelihood, \cite{Stoeber2013} proposed the following stepwise Expectation
Maximization (EM) algorithm based on the procedure of \cite{Aas2009}:
\begin{itemize}
\item [1.] Given the current parameters $(\bm{\theta}^{1,l}, \bm{\theta}^{2,l}, \bm{\mathrm{P}}^l)$ calculate iteratively the so-called \emph{\enquote{smoothed} probabilities}
\begin{eqnarray}
(\Omega_{t|T}((\bm{\theta}^{1,l}, \bm{\theta}^{2,l}, \bm{\mathrm{P}}^l))_{s_t}:=P(S_t=s_t |\bm{\mathrm{u}}_{1:T}, (\bm{\theta}^{1,l}, \bm{\theta}^{2,l}, \bm{\mathrm{P}}^l)),\quad s_t\in\{1,2\},
\end{eqnarray}
via a Hamilton filter (cf. \cite{Hamilton1989} and \cite{Stoeber2013} for details).
\item [2.] Maximize the pseudo log-likelihood function \small
\begin{eqnarray}\nonumber
&&Q((\bm{\theta}^{1,l+1}, \bm{\theta}^{2,l+1}, \bm{\mathrm{P}}^{l+1}); \bm{\mathrm{u}}_{1:T}, (\bm{\theta}^{1,l}, \bm{\theta}^{2,l}, \bm{\mathrm{P}}^l))\\\nonumber
&:=&\sum_{s_1=1}^2\ldots\sum_{s_T=1}^2 \log \left( f(\bm{\mathrm{u}}_{1:T}, s_{1:T} | (\bm{\theta}^{1,l+1}, \bm{\theta}^{2,l+1}, \bm{\mathrm{P}}^{l+1})) \right) P(S_{1:T}=s_{1:T}| \bm{\mathrm{u}}_{1:T},
(\bm{\theta}^{1,l}, \bm{\theta}^{2,l}, \bm{\mathrm{P}}^l)) \nonumber \\\nonumber
&\propto&\sum_{t=1}^T\left[\sum_{s_1=1}^2\ldots\sum_{s_T=1}^2 \log \left( f(\bm{\mathrm{u}}_{t},  | S_{t}=s_{t}, (\bm{\theta}^{1,l+1}, \bm{\theta}^{2,l+1})) \right) P(S_{1:T}=s_{1:T}| \bm{\mathrm{u}}_{1:T},
(\bm{\theta}^{1,l}, \bm{\theta}^{2,l}, \bm{\mathrm{P}}^l))\right]\nonumber \\\nonumber
&&+ \sum_{s_1=1}^2\ldots\sum_{s_T=1}^2\left[\sum_{t=1}^T \log \left( P(S_t=s_t | S_{t-1}, \bm{\mathrm{P}}^{l+1} )\right) + \log(P(S_1=s_1)^{l+1}) \right]\\ &&\quad\quad\quad\quad\quad\quad\times  P(S_{1:T}=s_{1:T}| \bm{\mathrm{u}}_{1:T},
(\bm{\theta}^{1,l}, \bm{\theta}^{2,l}, \bm{\mathrm{P}}^l))
\end{eqnarray}\normalsize 
stepwise, starting with respect to $(\bm{\theta}^{1,l+1}, \bm{\theta}^{2,l+1})$ which is carried out sequentially over the vine trees. Afterwards, the maximization with respect to $\bm{\mathrm{P}}^{l+1}$ can be done analytically via the formula of \cite{KimNelson}.
\end{itemize}
\quad\\

\section{Determining \enquote{normal} and \enquote{abnormal} regimes}\label{sec5}

The previous section outlined our MS-RV model setup which requires pre-defined $R$-vine structures for both regimes as input factors. Therefore, starting with the trees from model (2-dependent), we need to identify potential changes within the dependency structure of our index data over time. In order to do that, we shall follow \cite{Stoeber2013} and perform a detailed rolling window analysis (RWA) for the bivariate copulas within and between continents. Afterwards we start by estimating the MS-RV setup for each region separately, resulting in a global model assuming independent continents which we shall call (3-independent-MS). Finally, (3-dependent-MS) will be the corresponding Markov-switching setting for (2-dependent) which, up until now, performed best.
\quad\\

\subsection{Rolling window analysis}
For the following procedure, we choose 250-day windows into the future and consider which of our copulas out of Gauss, Student-$t$ and all Gumbel rotations fits best. Additionally, the value of Kendall's $\tau$ and the QTD from Definition~\ref{qtddef} is calculated as well. Therefore, in particular, we will be able to identify times of high and low tail dependence.

The outcome of the RWA is structured graphically as follows: For each index pair, the first graph shows the fitted copula family: Gauss (Ga), Student-$t$ (St-$t$), Gumbel (Gu), Gumbel90 (Gu90), Gumbel180 (Gu180) and Gumbel270 (Gu270). The second graph depicts the series of Kendall's $\tau$ based on \eqref{Taus} and as a horizontal line it's empirical value over the whole observation period. Finally, the third graph illustrates the QTDs and their mean over all calculcated values. The results for the first trees can be found in Figures~\ref{2ASIA_switching_1_col},~\ref{2EU_switching_1_col},~\ref{2USA_switching_1_col} and~\ref{2cont_switching_1_col}. These show that there are strong indications for regime shifts within the bivariate relationships of most index pairs. However, the plots for all higher order trees would not paint such a clear picture. In fact, the chosen copula families tend to vary more which can be explained by our earlier observations from Section~\ref{sec3} that the values of Kendall's $\tau$ for the second and third trees are generally much lower as for the first ones.

Considering the depicted results in more detail, we realize that especially for the respective equity-volatility pairs, Gumbel90 seems to be the standard regime indicating a nearly constant asymmetric tail dependence which is further supported by the QTD values. In the case of an equity-commodity relation, i.e. SPX-BBC, Kendall's $\tau$ as well as the QTD values increased rather sharply in 2008 and started to came back to their previous levels near zero 2014. This clearly illustrates the vanishing diversification in a global crisis, as the strength of positive dependence between equity and commodities tends to increase during bad market times.

\begin{sidewaystable}[htbp]
\centering
\begin{tabular}{c c  l  }
& & \hspace{2cm} Copula Family \hspace{4cm} Kendall's Tau \hspace{3.3cm} Quarter Tail Dependence  \\
\rotatebox{90}{\hspace{1.5cm} Eq-Eq}&  \rotatebox{90}{\hspace{1.2cm} NKY-HSI}&  \includegraphics[height=4cm]{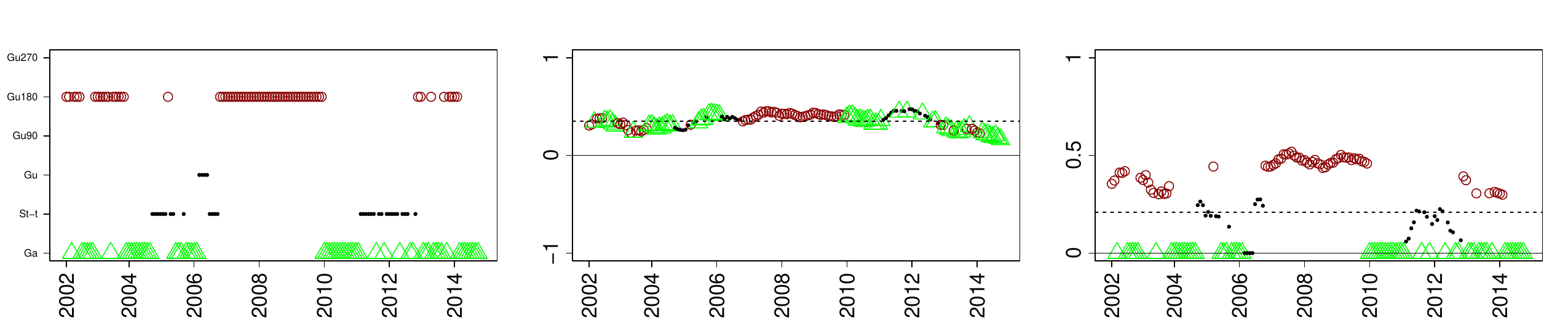}  \\
\rotatebox{90}{\hspace{1.5cm} Eq-Vol}&         \rotatebox{90}{\hspace{1cm} NKY-VNKY}&  \includegraphics[height=4cm]{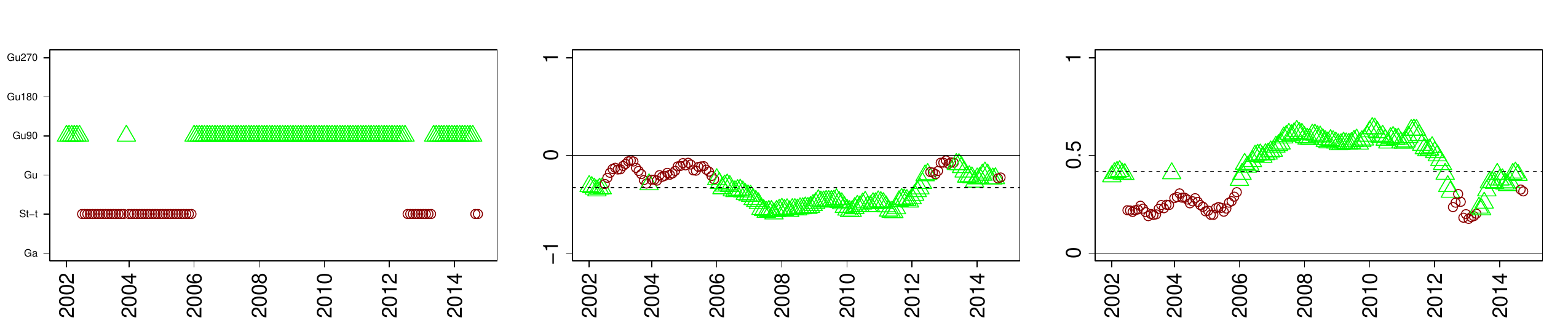}  \\
\rotatebox{90}{\hspace{1.5cm} Eq-Vol}&               \rotatebox{90}{\hspace{1.2cm} HSI-VHSI} &  \includegraphics[height=4cm]{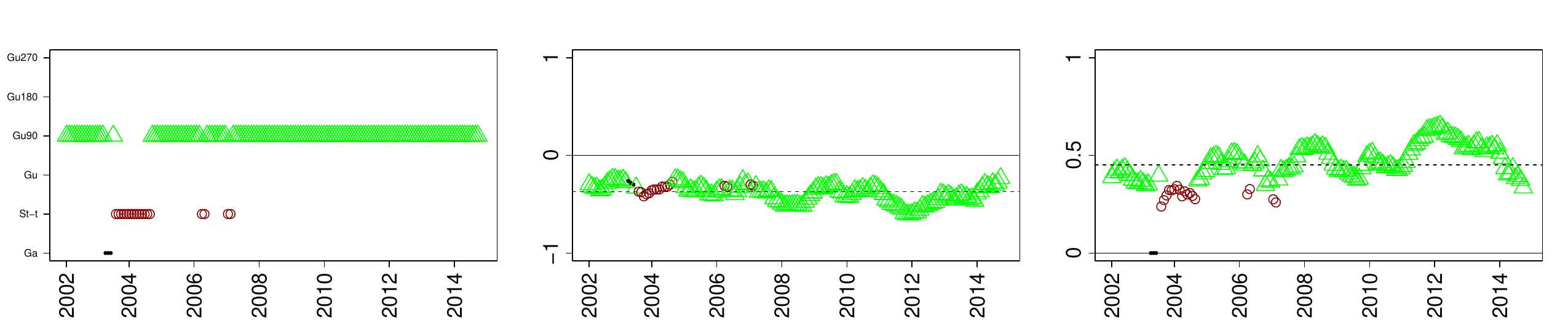}  \\
\end{tabular}
\captionof{figure}{Asia - 250 day rolling window analysis for bivariate copulas in the first tree. Periods are marked in green/triangle and red/star, respectively, when copula families chosen to represent \enquote{normal} and \enquote{abnormal} regimes were fitted. For reasons of visibility, only every 20th data point is plotted.}
\label{2ASIA_switching_1_col}
\end{sidewaystable}

\begin{sidewaystable}[htbp]
\centering
\begin{tabular}{c c  l  }
& & \hspace{2cm} Copula Family \hspace{4cm} Kendall's Tau \hspace{3.3cm} Quarter Tail Dependence  \\
\rotatebox{90}{\hspace{1.5cm} Eq-Eq}&  \rotatebox{90}{\hspace{1cm} DAX-SX5E}&  \includegraphics[height=4cm]{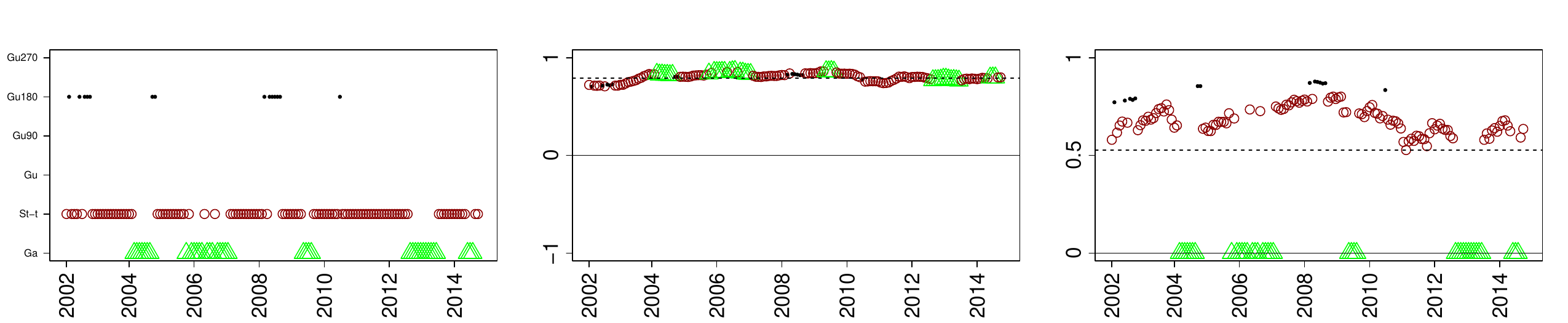}  \\
\rotatebox{90}{\hspace{1.5cm} Eq-Vol}&        \rotatebox{90}{\hspace{1cm} DAX-VDAX}&  \includegraphics[height=4cm]{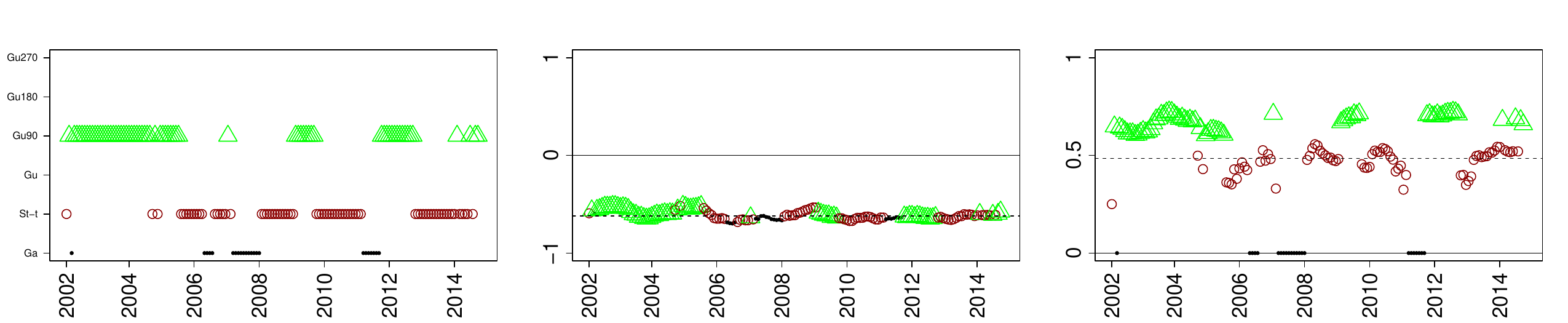}  \\
\rotatebox{90}{\hspace{1.5cm} Eq-Vol}&              \rotatebox{90}{\hspace{1cm} SX5E-VSX5E}&  \includegraphics[height=4cm]{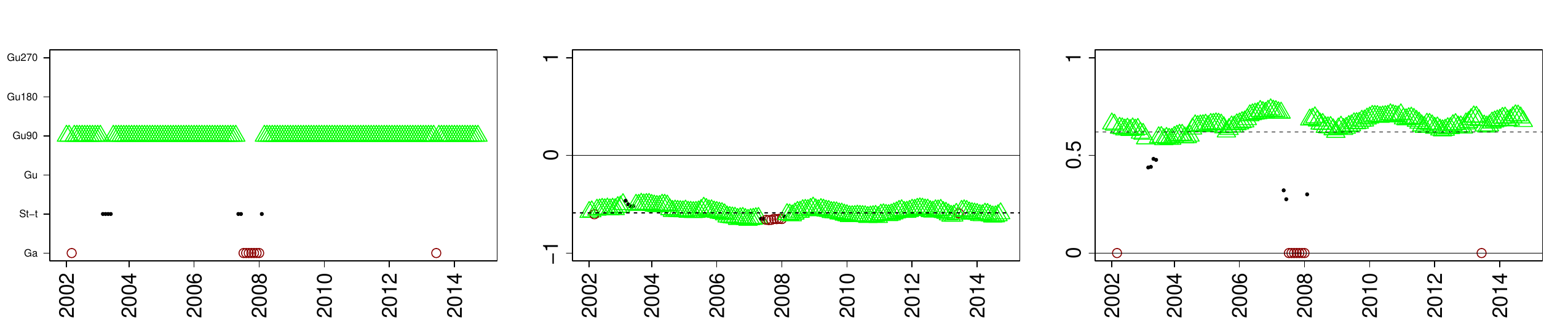}  \\
\end{tabular}
\captionof{figure}{Europe - 250 day rolling window analysis for bivariate copulas in the first tree. Periods are marked in green/triangle and red/star, respectively, when copula families chosen to represent \enquote{normal} and \enquote{abnormal} regimes were fitted. For reasons of visibility, only every 20th data point is plotted.}
\label{2EU_switching_1_col}
\end{sidewaystable}

\begin{sidewaystable}[htbp]
\centering
\begin{tabular}{c c  l  }
& & \hspace{2cm} Copula Family \hspace{4cm} Kendall's Tau \hspace{3.3cm} Quarter Tail Dependence  \\
\rotatebox{90}{\hspace{1.5cm} Eq-Vol}&  \rotatebox{90}{\hspace{1.2cm} SPX-VIX}&  \includegraphics[height=4cm]{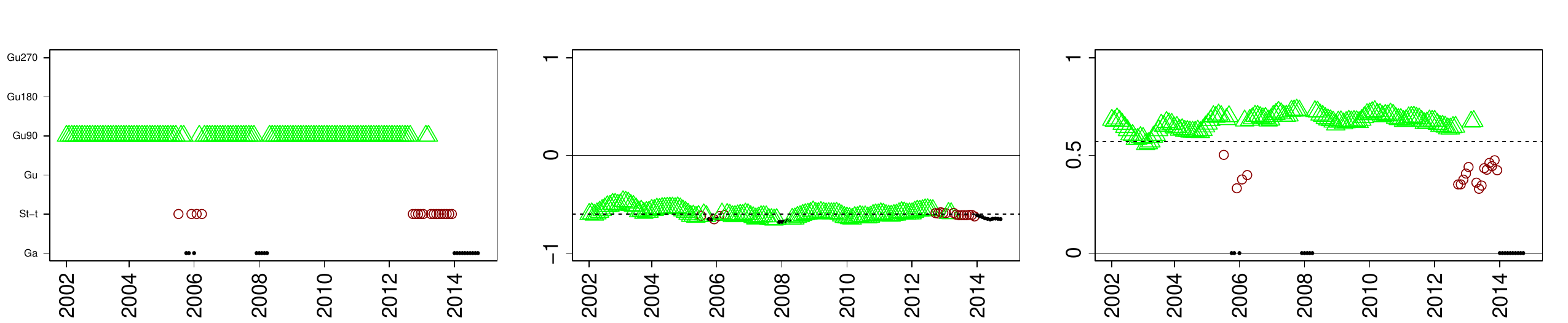}  \\
\rotatebox{90}{\hspace{1.5cm} Eq-Cmd}&        \rotatebox{90}{\hspace{1.3cm} SPX-BBC}&  \includegraphics[height=4cm]{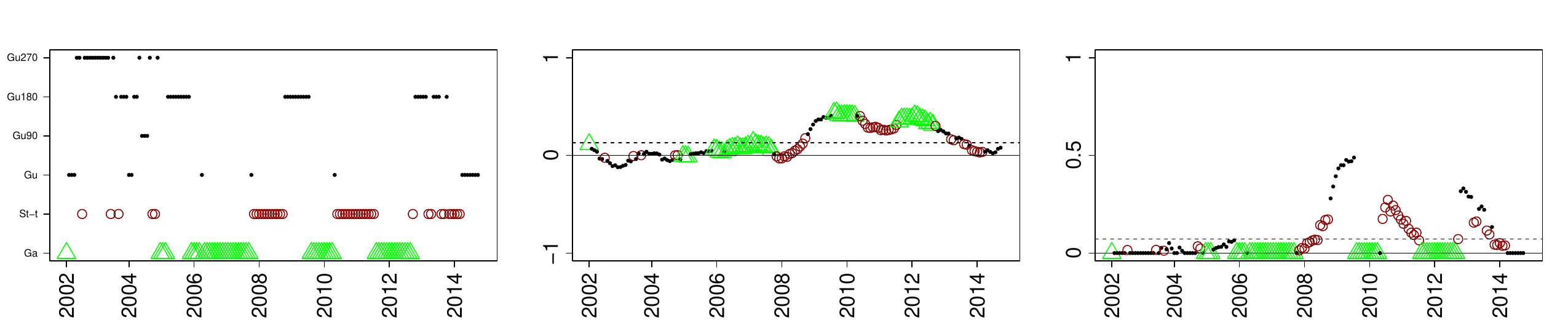}  \\
\end{tabular}
\captionof{figure}{USA - 250 day rolling window analysis for bivariate copulas in the first tree. Periods are marked in green/triangle and red/star, respectively, when copula families chosen to represent \enquote{normal} and \enquote{abnormal} regimes were fitted. For reasons of visibility, only every 20th data point is plotted.}
\label{2USA_switching_1_col}
\end{sidewaystable}

\begin{sidewaystable}[htbp]
\centering
\begin{tabular}{c c  l  }
& & \hspace{2cm} Copula Family \hspace{4cm} Kendall's Tau \hspace{3.3cm} Quarter Tail Dependence  \\
\rotatebox{90}{\hspace{1.5cm} Eq-Eq}&  \rotatebox{90}{\hspace{1.2cm} HSI-SX5E}&  \includegraphics[height=4cm]{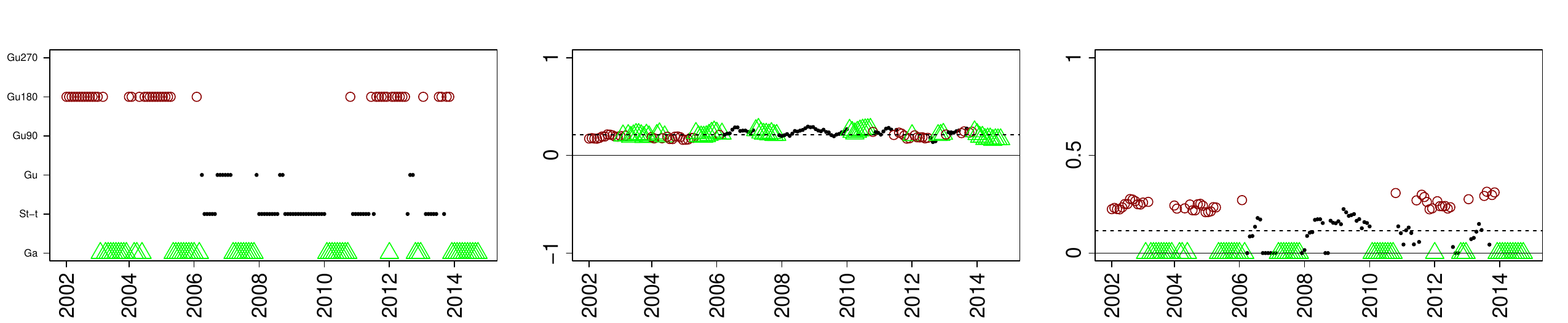}  \\
\rotatebox{90}{\hspace{1.5cm} Eq-Eq}&        \rotatebox{90}{\hspace{1.3cm} DAX-SPX}&  \includegraphics[height=4cm]{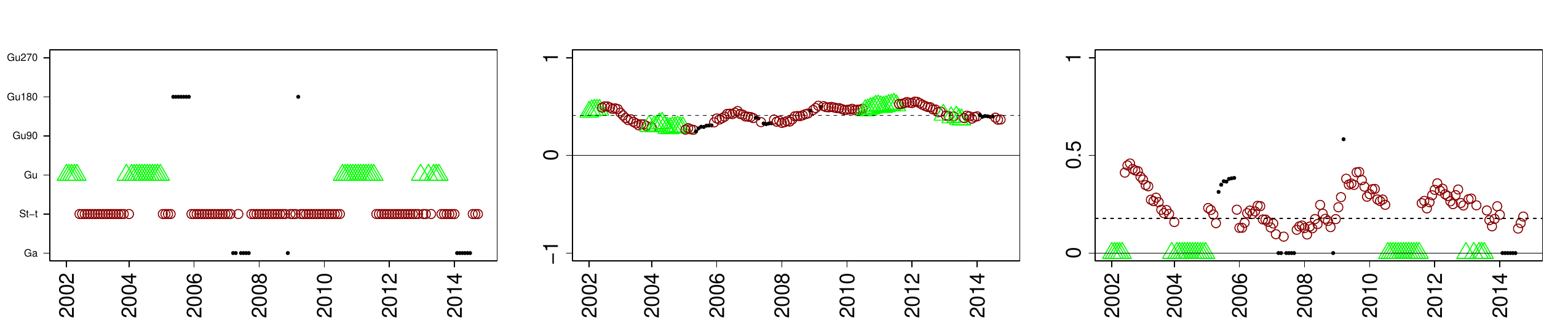}  \\
\end{tabular}
\captionof{figure}{Between continents - 250 day rolling window analysis for bivariate copulas in the first tree. Periods are marked in green/triangle and red/star, respectively, when copula families chosen to represent \enquote{normal} and \enquote{abnormal} regimes were fitted. For reasons of visibility, only every 20th data point is plotted.}
\label{2cont_switching_1_col}
\end{sidewaystable}

\begin{figure}[htbp]
\centering
\includegraphics[height=5cm]{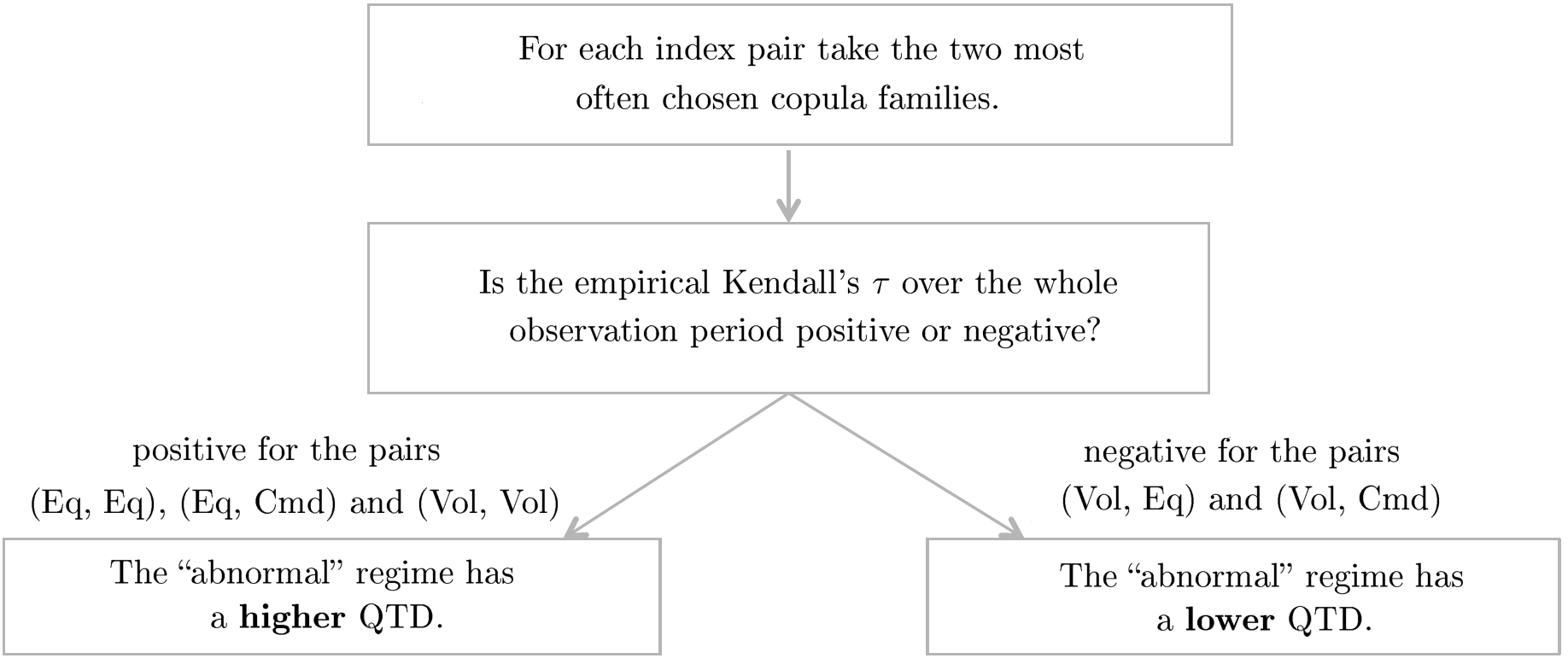}
\caption{Classification algorithm for determining \enquote{normal} and \enquote{abnormal} regimes within the first trees based on the (theoretical) Kendall's $\tau$ and QTD.}
\label{Algorithm}
\end{figure}

The equity-equity index pairs on the flipside do not show such a clear picture: While Kendall's $\tau$ is comparably high for DAX-SX5E and DAX-SPX and the Gauss/Student-$t$ copulas are predominantly chosen, the connection between the developed regions and China given by HSI-SX5E as well as NKY-HSI seems weaker and tends to either a Gaussian or Gumbel180 copula. However, the QTD values are generally higher in abnormal times like 2007-2010 or 2010-2012 - the only exception being DAX-SX5E which shows quite a strong QTD for nearly the whole period. Anyhow, this can be explained by the fact that the DAX is partly included within the SX5E.

Having considered the RWA results as a kind of explanatory analysis, we have to specify how \enquote{normal} and \enquote{abnormal} shall be classified within our Markov-switching setup, i.e. what kind of behavior in the dependence structure of our indices is usually not expected in \enquote{normal} times. For potential regime switches within the first trees, we propose the algorithm depicted by Figure~\ref{Algorithm}: Firstly, the two copula families that appear most often within the RWA are identified for each index pair. Secondly, we differentiate between pairs with overall positive and negative dependence measured by the empirical Kendall's $\tau$. In case of positive dependence, the \enquote{abnormal} regime should be the one with the higher QTD among the two, which is inline with our findings above. For negative values of the empirical Kendall's $\tau$, as can be found for, e.g., equity-volatility pairs, the \enquote{abnormal} state is identified as the one with the lower QTD corresponding to the fact that Gumbel90 has been shown to be the \enquote{normal} regime. For all trees of higher orders, justified by the rather low values of Kendall's $\tau$, we shall fix the copula family with the highest frequency in the RWA and allow only the individual parameters to switch between regimes. The final classifications are summarized by Table~\ref{regimes}.

As in \cite{Stoeber2013}, before running the EM algorithm from Section~\ref{sec4} to estimate the MS-RV model, we want to check if our chosen regimes are sensible by considering the differences in the log-likelihoods over time. For that, both copula families for the \enquote{normal} and \enquote{abnormal} regimes in the first tree are fitted separately using again 250-day rolling windows. Figure~\ref{ll_difference} summarizes the results by plotting the difference
\begin{equation} \label{loglik_diff}
\log L(C_{abnormal})-\log L(C_{normal}),
\end{equation}
i.e. positive values indicate an outperformance of the \enquote{abnormal} regime. Additionally, the periods are marked, when the copula families identified with \enquote{normal} and \enquote{abnormal} regimes were fitted in the RWA. The plotted results confirm that our algorithm (cf. Figure~\ref{Algorithm}) seems a sensible choice.

\begin{sidewaystable}
\centering
\begin{tabular}{l l | c | c c c | c c c}
& & &\multicolumn{3}{c|}{\enquote{Normal} regime} & \multicolumn{3}{c}{\enquote{Abnormal} regime}\\
\multicolumn{2}{c|}{Index Pair	}& Kendall's $\tau$ &Family & Frequency & $\lambda_{QTD}$ & Family & Frequency & $\lambda_{QTD}$ \\ \hline \hline
\multicolumn{9}{c}{} \\
\multicolumn{9}{c}{Asia} \\\hline
Eq-Eq &NKY-HSI & $>0$ & \textbf{Gauss} & 0.39 & 0 & \textbf{Gumbel180} & 0.41 & 0.43\\
Eq-Vol &NKY-VNKY &$<0$& \textbf{Gumbel90} & 0.67 &0.51 & \textbf{Student-$t$} &0.33 & 0.24\\
Eq-Vol &HSI-VHSI &$<0$& \textbf{Gumbel90} & 0.88 &0.48 &\textbf{Student-$t$} &0.11 & 0.30\\
Eq-Vol &HSI-VNKY$|$NKY  &$<0$& \textbf{Gumbel90} & 0.48 &0.10 & \textbf{Gumbel90} & 0.48 &0.10\\
Eq-Vol &NKY-VHSI$|$HSI  &$<0$&\textbf{Gauss} &0.37 & 0 &\textbf{Gauss} &0.37 & 0\\
Vol-Vol &VNKY-VHSI$|$NKY, HSI  & $>0$ & \textbf{Gauss} & 0.49 &0 & \textbf{Gauss} & 0.49 &0\\ \hline
\multicolumn{9}{ c}{} \\
\multicolumn{9}{ c}{Europe} \\\hline
Eq-Eq &DAX-SX5E & $>0$ & \textbf{Gauss} & 0.24 &0 &  \textbf{Student-$t$} & 0.69 & 0.68\\
Eq-Vol &DAX-VDAX &$<0$& \textbf{Gumbel90} & 0.42 &0.67 & \textbf{Student-$t$} & 0.43 & 0.47 \\
Eq-Vol &SX5E-VSX5E &$<0$& \textbf{Gumbel90} & 0.91 &0.66 &\textbf{Gauss}& 0.04 & 0\\
Eq-Vol &SX5E-VDAX$|$DAX &$<0$&\textbf{Gauss}& 0.40 & 0 &\textbf{Gauss}& 0.40 & 0\\
Eq-Vol & DAX-VSX5E$|$SX5E &$<0$&\textbf{Gumbel270}&0.31 & 0 &\textbf{Gumbel270}&0.31 & 0\\
Vol-Vol & VDAX-VSX5E$|$DAX,SX5E & $>0$ &  \textbf{Student-$t$}& 0.66 & 0.33 &  \textbf{Student-$t$}& 0.66 & 0.33\\ \hline
\multicolumn{9}{ c}{} \\
\multicolumn{9}{ c}{USA} \\\hline
Eq-Vol &SPX-VIX &$<0$& \textbf{Gumbel90} & 0.78 &0.67 &   \textbf{Student-$t$}& 0.12 & 0.40 \\
Eq-Cmd & SPX-BBC & $>0$ & \textbf{Gauss} & 0.32 &0 &   \textbf{Student-$t$}&0.25 & 0.10\\
Vol-Cmd & VIX-BBC$|$SPX &$<0$ & \textbf{Gauss}&0.38 & 0 & \textbf{Gauss}&0.38 & 0\\ \hline
\multicolumn{9}{ c}{} \\
\multicolumn{9}{ c}{Between continents} \\\hline
Eq-Eq&HSI-SX5E & $>0$ & \textbf{Gauss} & 0.34 &0 &\textbf{Gumbel180}& 0.32 & 0.25\\
Eq-Eq & DAX-SPX & $>0$ & \textbf{Gumbel} & 0.22 &0 & \textbf{Student-$t$}& 0.62 & 0.25\\ \hline
\end{tabular}
\caption{Results of our algorithm (Figure~\ref{Algorithm}): \enquote{Normal} and \enquote{abnormal} regime classification by frequency and quarter tail dependence $\lambda_{QTD}$ for the first trees. For all higher trees the copulas with the highest frequencies in the RWA are chosen and only the individual parameters are allowed to switch.}
\label{regimes}
\end{sidewaystable}

\begin{sidewaystable}[]
\centering
\begin{tabular}{c |c | c | c }
 \multicolumn{4}{c}{Difference of the log-likelihoods of the fit of two copula families: "abnormal" and "normal"}\\ \multicolumn{4}{c}{}\\
 \textbf{Asia} & \textbf{Europe} & \textbf{North-America} & \textbf{Between continents}\\
   \subfloat[NKY-HSI]{\includegraphics[height=3.0cm]{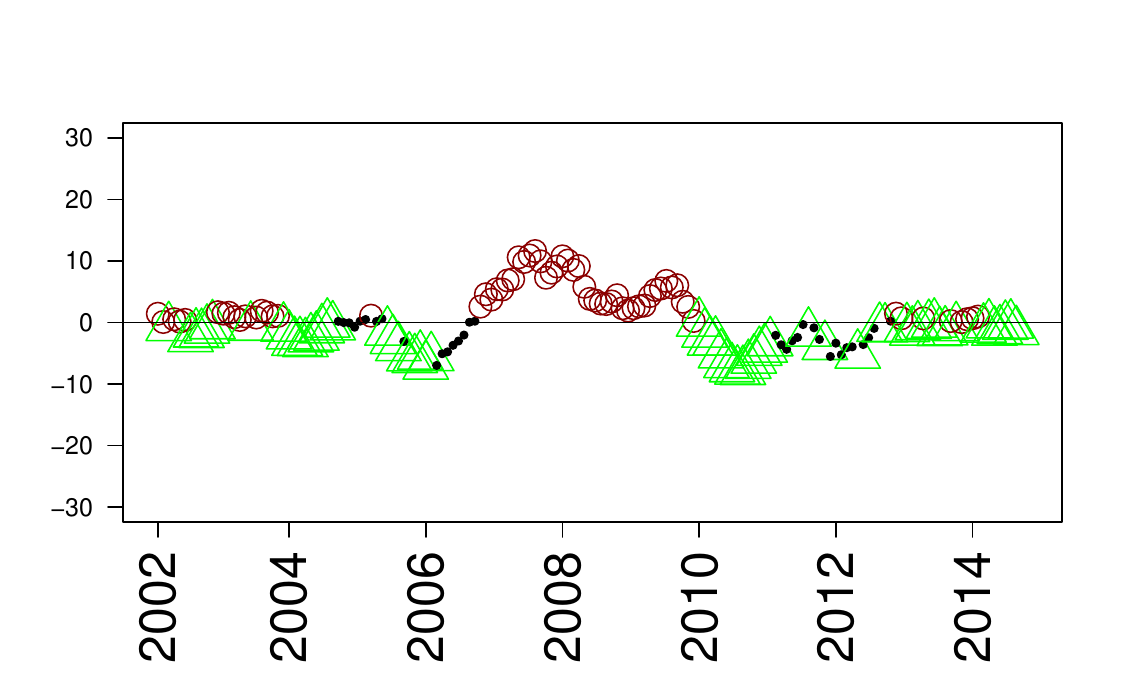}}
  & \subfloat[DAX-SX5E]{\includegraphics[height=3.0cm]{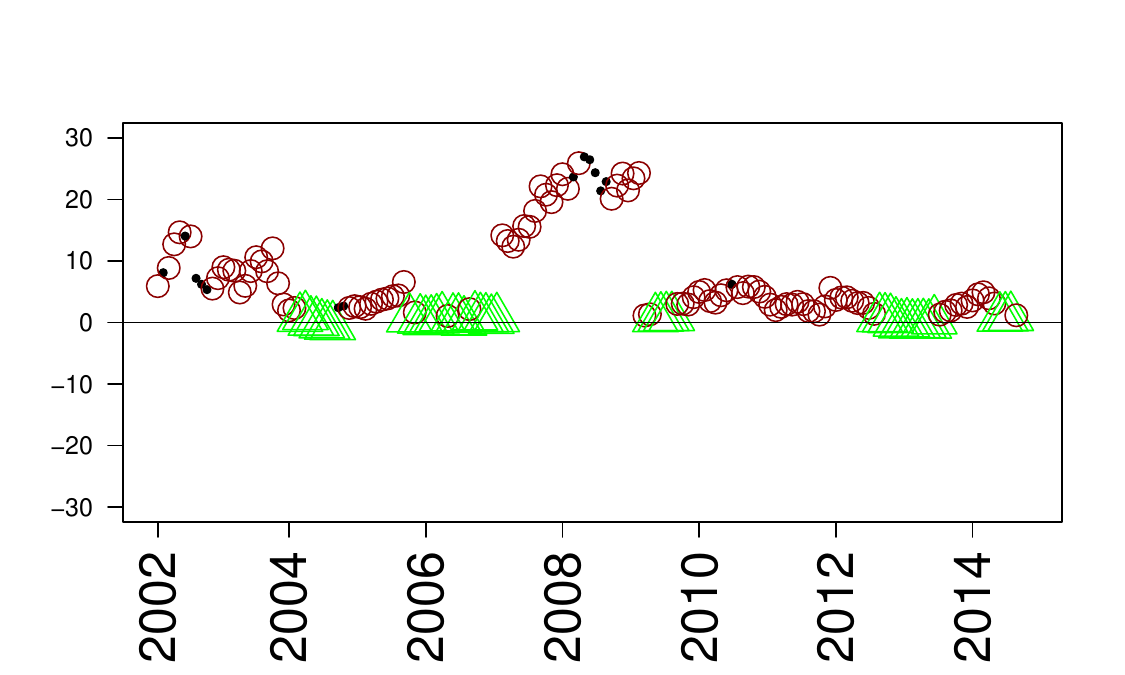}}
  &\subfloat[SPX-VIX]{ \includegraphics[height=3.0cm]{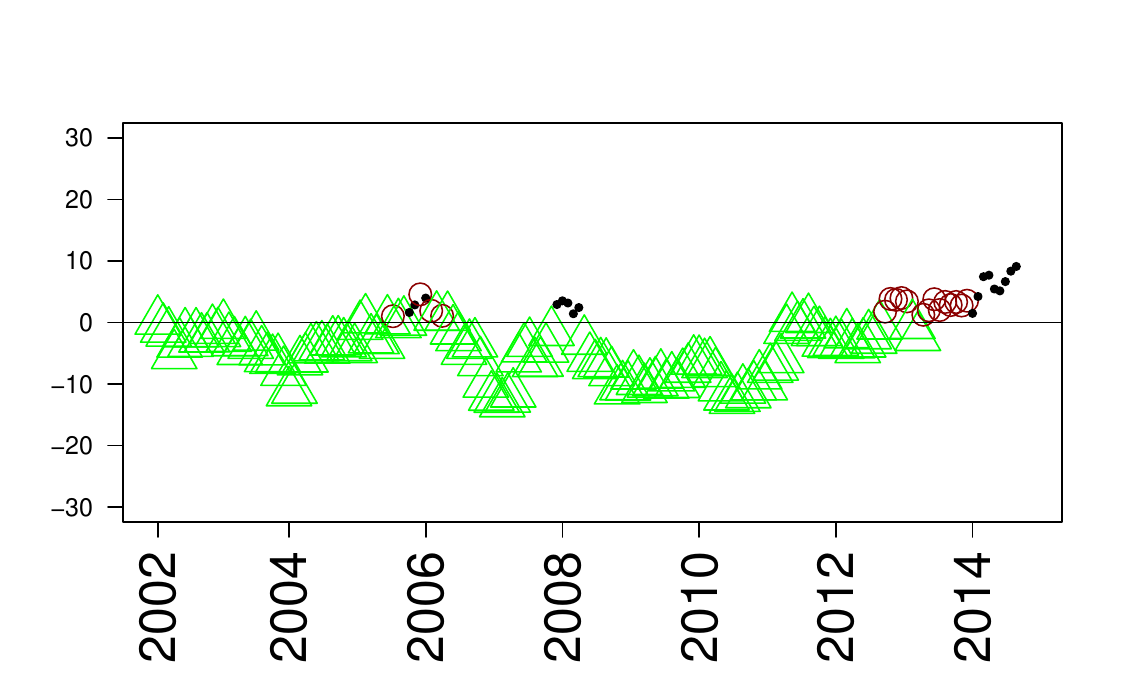}} & \subfloat[HSI-SX5E]{\includegraphics[height=3.0cm]{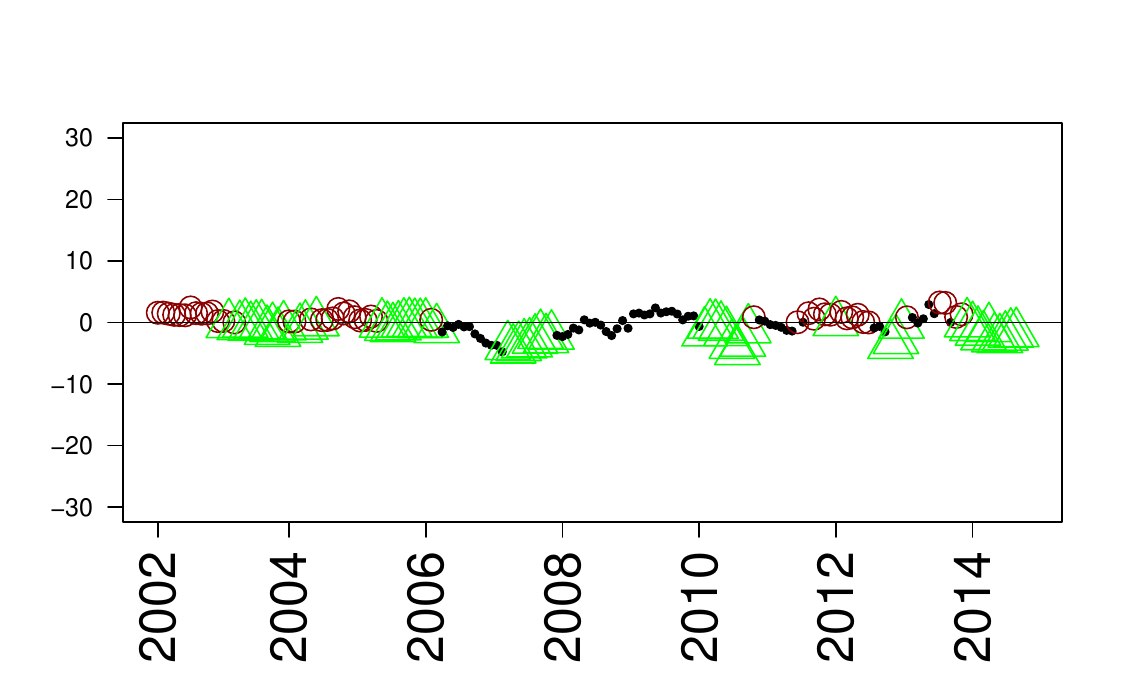}}\\
     \subfloat[NKY-VNKY]{\includegraphics[height=3.0cm]{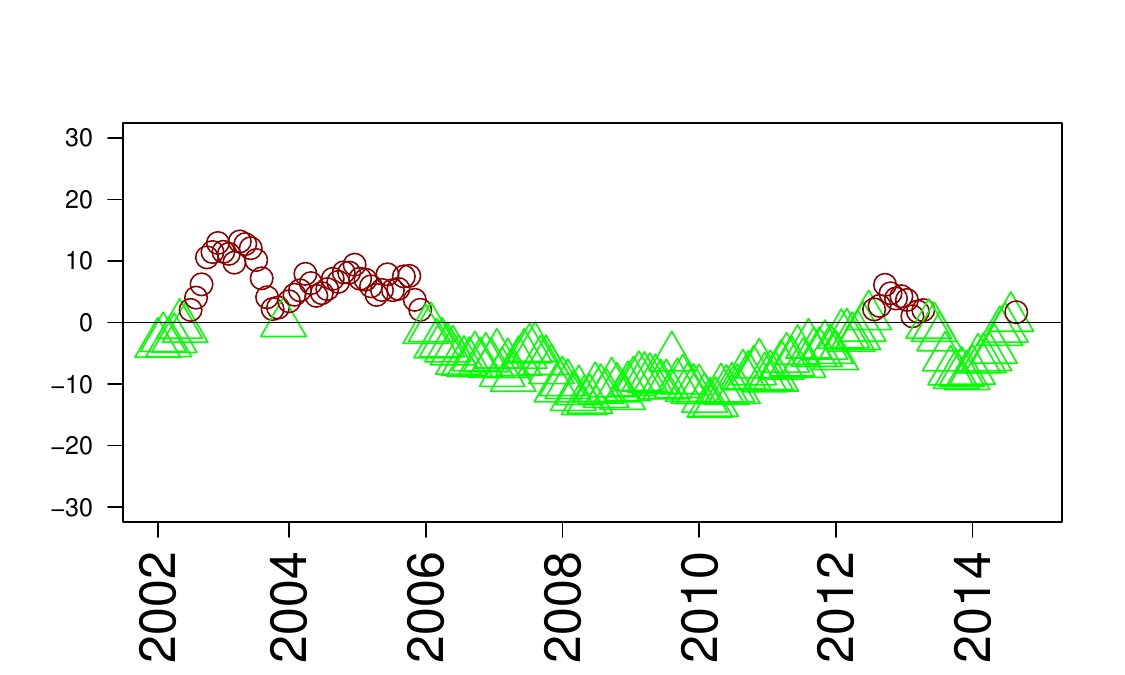}}
     & \subfloat[DAX-VDAX]{\includegraphics[height=3.0cm]{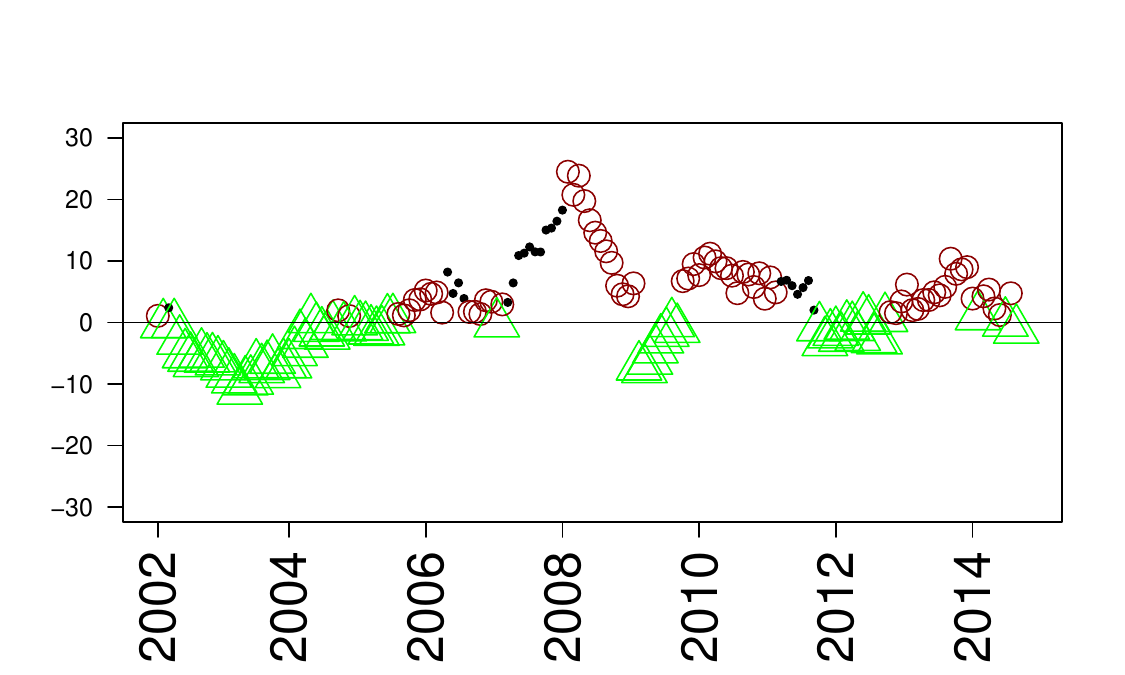}}
     & \subfloat[SPX-BBC]{\includegraphics[height=3.0cm]{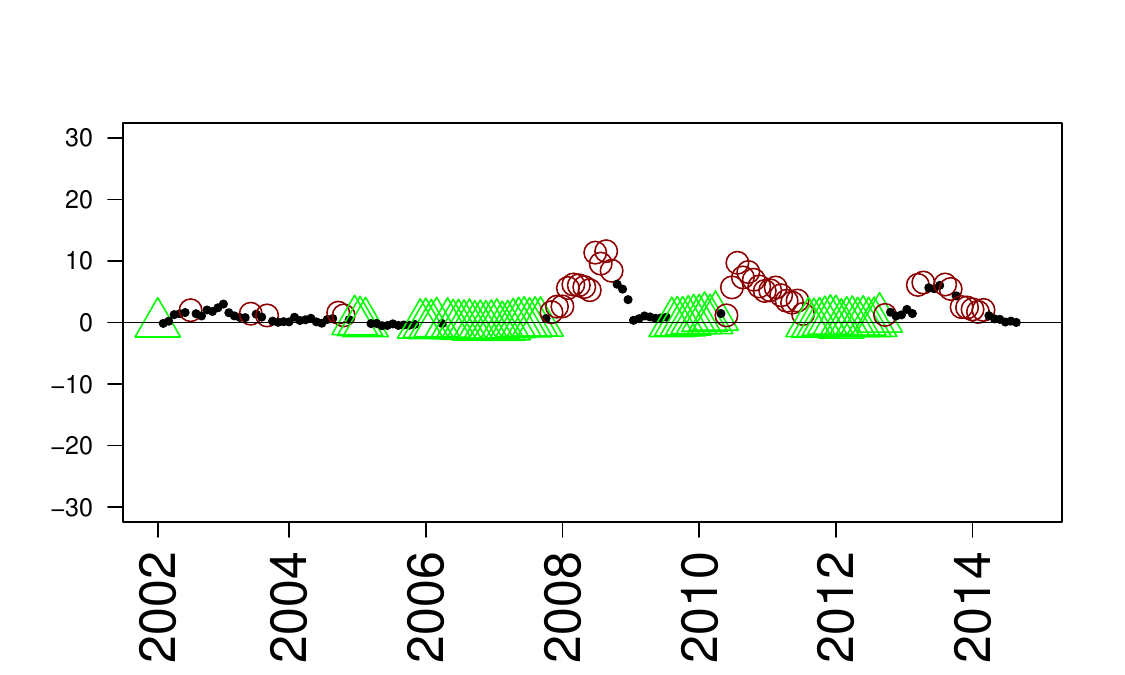}} & \subfloat[DAX-SPX]{\includegraphics[height=3.0cm]{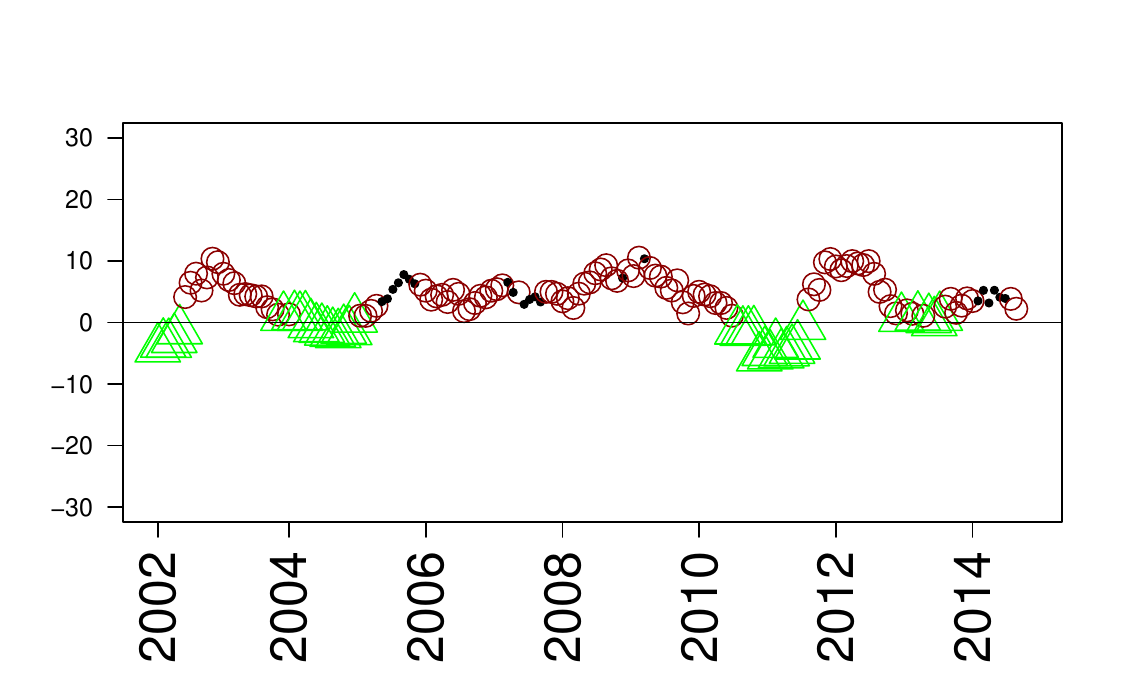}}\\
            \subfloat[HSI-VHSI]{\includegraphics[height=3.0cm]{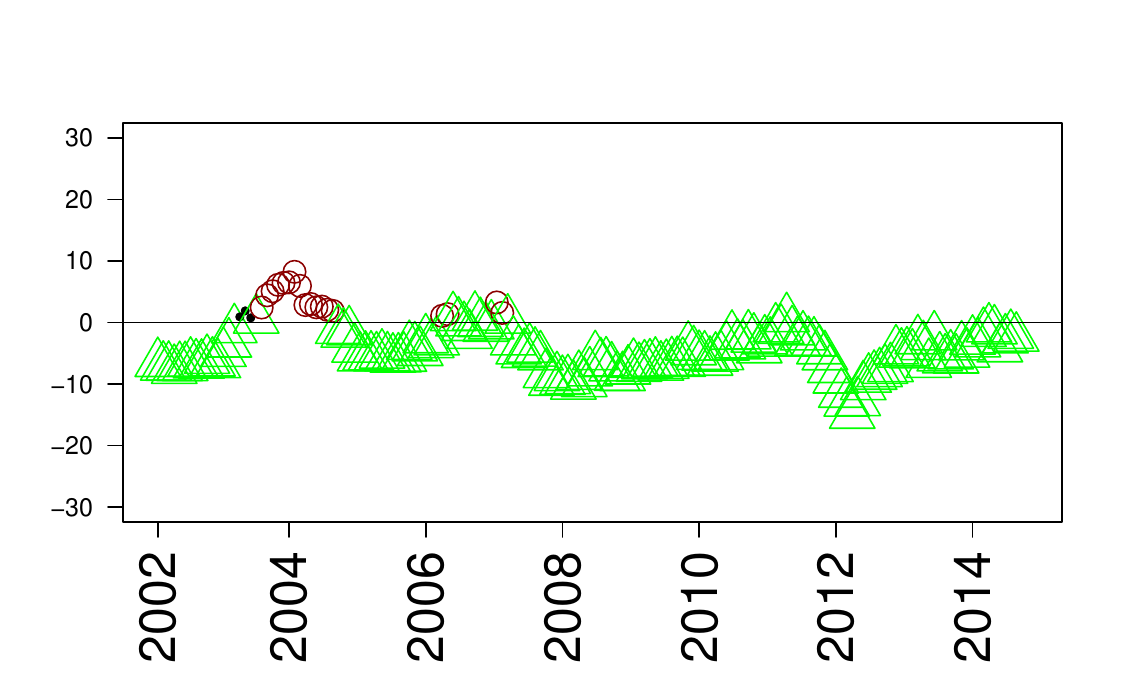}}
           & \subfloat[SX5E-VSX5E]{\includegraphics[height=3.0cm]{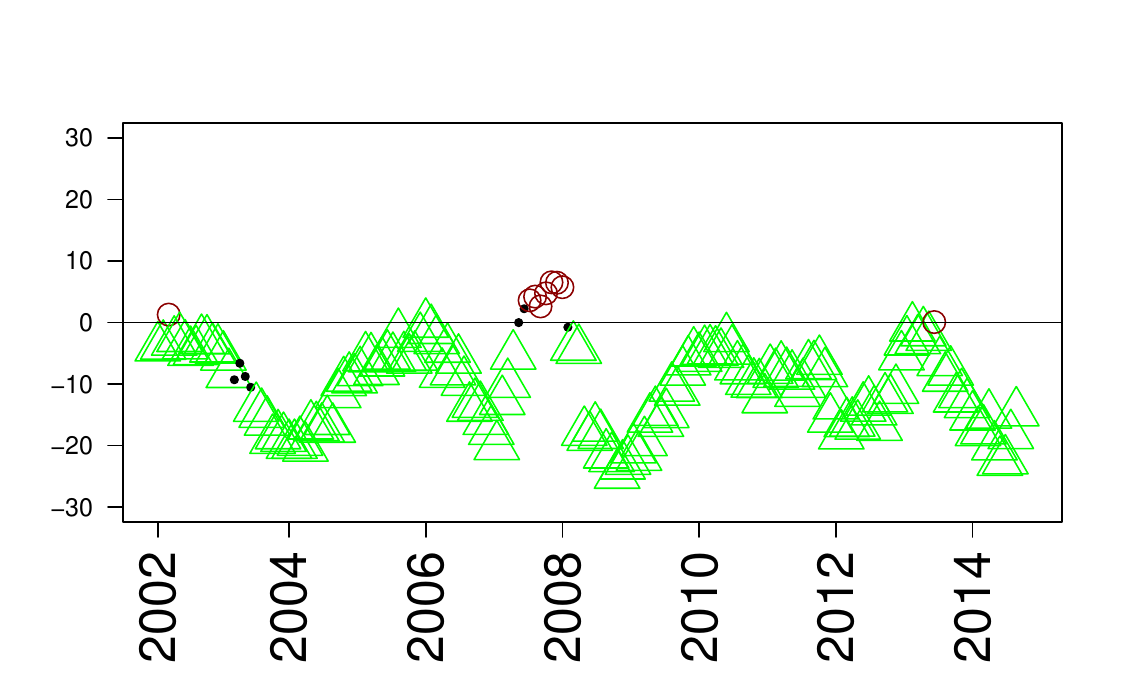}}
           & &
   \end{tabular}
	\captionof{figure}{Difference in the log-likelihoods between the fit of the copula family chosen to represent \enquote{abnormal} and \enquote{normal} regime. Positive values indicate outperformance of the \enquote{abnormal} regime. Again, periods are marked as in Figures~\ref{2ASIA_switching_1_col}, \ref{2EU_switching_1_col}, \ref{2USA_switching_1_col} and \ref{2cont_switching_1_col}.}
	\label{ll_difference}
\end{sidewaystable}

\begin{table}[h]
\centering
\begin{tabular}{ l | r | r | r }
& Log-likelihood &\multicolumn{1}{ c| }{AIC}  &\multicolumn{1}{c }{BIC} \\ \hline \hline
Model Asia$_1$ 		&2,027.61  &-4,041.22 &-3,998.24\\
Model Asia$_{1,\mathbf{MS}}$	& \textbf{2,118.59} &\textbf{-4205.18}& \textbf{-4,106.92}\\ \hline
Model Europe$_1$ 		&8,973.48  &-17,924.95 &-17,857.40\\
Model Europe$_{1,\mathbf{MS}}$		&  \textbf{9,053.72}& \textbf{-18,071.44}& \textbf{-17,960.89}\\ \hline
Model USA$_1$ 		& 2,017.63&  -4,027.26 & -4,002.69\\
Model USA$_{1,\mathbf{MS}}$ 		& \textbf{2,074.60}&  \textbf{-4,135.20} & \textbf{-4,092.22}\\ \hline
\end{tabular}
\caption{Log-likelihood, AIC and BIC for the static $R$-vine models Asia$_1$, Europe$_1$, USA$_1$ and their Markov-switching counterparts.}
\label{model_comparison_switch}
\end{table}

\begin{table}[h]
\centering
\begin{tabular}{ l | r | r | r }
& Log-likelihood &\multicolumn{1}{ c| }{AIC}  &\multicolumn{1}{c }{BIC} \\ \hline \hline
(2-independent) 		&  13,018.72&  -25,993.44 & -25,858.33\\
(3-independent-MS) 		& \textbf{13,246.91}& \textbf{-26,411.81}& \textbf{-26,160.02}\\ \hline
(2-dependent) 		&  14,618.70& -29,091.40& \textbf{-28,643.09}\\
(3-dependent-MS) 		& \textbf{14,738.88}& \textbf{-29,191.76}& -28,313.56\\ \hline
\end{tabular}
\caption{Log-likelihood, AIC and BIC for the global, static $R$-vine models (2-independent) and (2-dependent) compared to their Markov-switching counterparts (3-independent-MS) and (3-dependent-MS).}
\label{model_comparison_all_switch}
\end{table}
\quad\\

\subsection{Results on MS-RV models}

\begin{figure}[htbp]
\centering
\includegraphics[height=15.5cm]{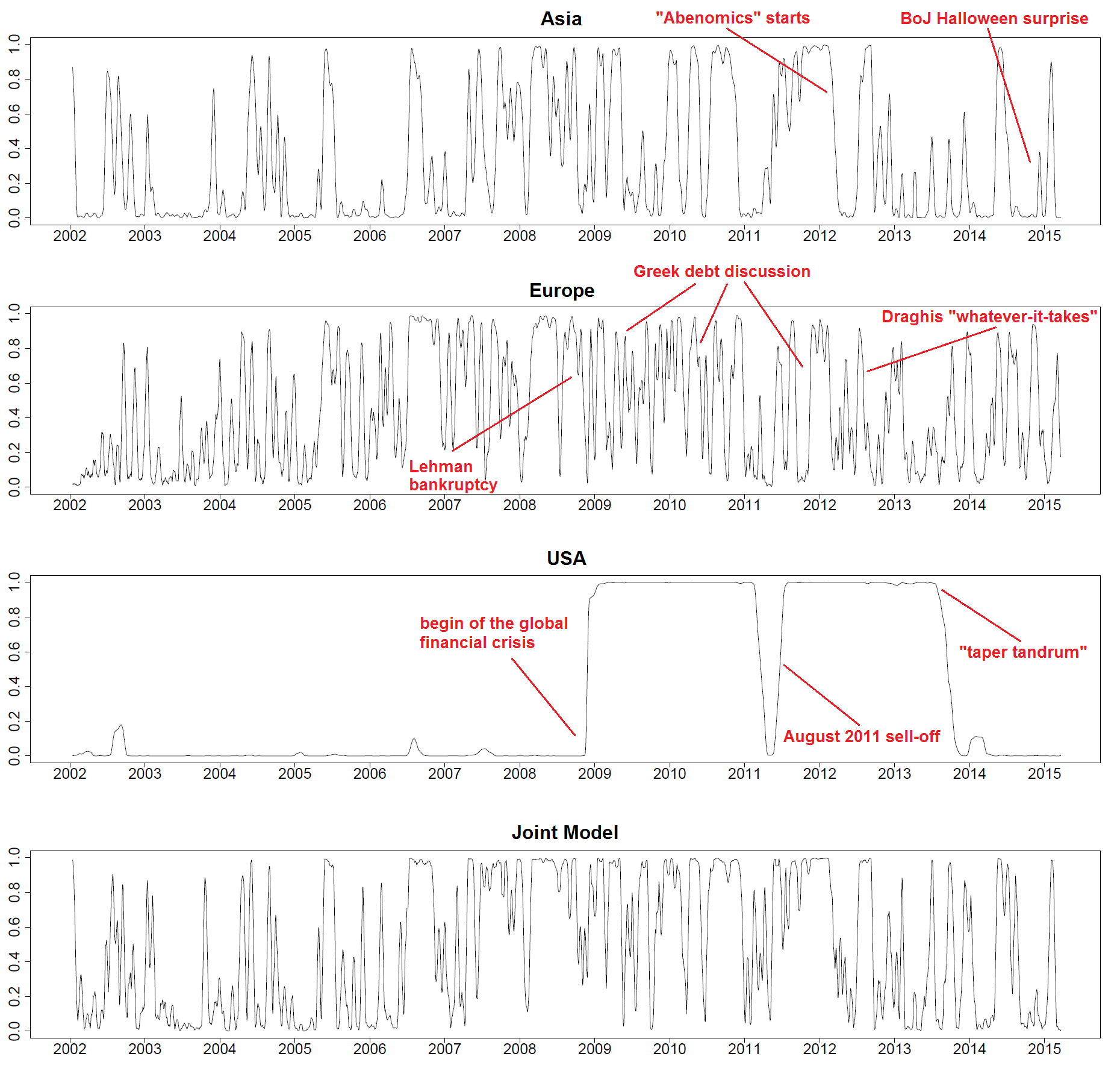}
\caption{Probabilities of being in the \enquote{abnormal} regime smoothed by 10-day moving average for model (3-independent-MS) and (3-dependent-MS).}
\label{pbb_crisis_cont_MA30}
\end{figure}

We start by considering each region individually again. The first trees were taken as determined in Table~\ref{regimes} based on the algorithm from Figure~\ref{Algorithm} while all higher trees are chosen inline with (2-independent), letting only the copula parameters switch. For all three continents, compared with their static counterparts, the MS-RV model performs better in terms of log-likelihood, AIC and BIC as outlined by Table~\ref{model_comparison_switch}. The probabilities of being in the \enquote{abnormal} regime, smoothed by a 10-day moving average, can be seen in the first three plots of Figure~\ref{pbb_crisis_cont_MA30}.

In the US we see a strong predominance of the \enquote{normal} regime pre-Lehman while after the start of the global financial crisis and up until the summer of 2013 the \enquote{abnormal} regime prevails. Having the results of the RWA in mind (cf. Figure~\ref{2USA_switching_1_col}), the regime switch to \enquote{abnormal} seems to be mainly driven by SPX-BBC as the general dependence structure of SPX-VIX is rather stable over time. In June 2013, the Federal Reserve announced the potential scaling back of its quantitative easing programm later that year leading to a more than 4\% fall in the equity markets and a spike in volatility (\enquote{taper tandrum}). Interestingly, this regime change can not be seen directly in the first tree and is merely driven by the parameter switches in the higher trees. The drop in the probability in early 2011 goes hand-in-hand with the fall in the QTD of SPX-BBC while the rise in mid-2011 can be explained by the equity market sell-off and spike in implied volatility in August 2011 which is again better reflected in the higher order trees.

Compared to the US, the plots for Asia and Europe clearly show a greater activity in terms of regime switching which mostly seems to happen at the same time indicating that a joint MS-RV model might be a sensible choice.

We want to stress again that by definition of our \enquote{abnormal} regimes, these do not necessarily indicate a crisis in financial markets. For example, the Halloween surprise by the Bank of Japan on 31 October 2014 when Governor Haruhiko Kuroda unexpectedly increased his QQE program by an annual pace of 30 trillion JPY is obviously not a crisis but can still be matched with a rising probability of being in the \enquote{abnormal} regime.
A similar feature can be seen in Japan throughout the year 2013: Due to the election of the Liberal Democrats under the leadership of Shinzo Abe in December 2012 and the subsequent implementation of \enquote{Abenomics}, both, the equity markets and implied volatility started to rise indicating an \enquote{abnormal} regime (which can also be seen in the RWA). However in Hong Kong, the HSI had a rather bad year (cf. Figure~\ref{levels}) with a rising VHSI drawing the probability plot down to the \enquote{normal} state again and again.

On the other side, the rather high frequency of switches in Europe especially in 2009-2012 can be explained by the ongoing Greek debt discussion (i.e. a classical crisis) ending with Mario Draghis famous \enquote{whatever it takes}-speech on 26 July 2012. What might be surprising on a first view is that the Lehman crisis brought the probability plot down and close to zero in the subsequent months. However, this can again be explained by the \enquote{normal} regime's definition and a look at the RWA: While for the equity markets, the \enquote{abnormal} regime might be sensible, the historically large rise in implied volatility in both, the VDAX and the VSX5E (cf. Figure~\ref{levels}), occurring at the same time draws our model back to the \enquote{normal} state which means a Gumbel90 copula for DAX-VDAX and SX5E-VSX5E.

Assuming independent continents and adding the log-likelihoods up, unsurprisingly shows that the above setup, called (3-independent-MS) outperforms its benchmark (2-independent), cf. Table~\ref{model_comparison_all_switch}. However, as we noted especially for Asia and Europe, switches seem to occur mostly at the same time, even though the special behavior of the USA would not directly support that hypothesis. Therefore as a final investigation, we estimate a joint MS-RV model, implying regime changes at the same time, which we shall call (3-dependent-MS). Again, the first trees are based on Table~\ref{regimes} while the higher order dependencies were chosen inline with (2-dependent) allowing only the copula parameters to switch. In terms of log-likelihood and AIC, the MS-RV model is superior to the static setup while BIC (which penalizes the amount of parameters more than AIC) rather supports the later one, cf. Table~\ref{model_comparison_all_switch}. The smoothed probability plot of model (3-dependent-MS) can be found at the bottom of Figure~\ref{pbb_crisis_cont_MA30}. Most of the previously identified periods of the \enquote{abnormal} regime appear here as well.
\quad\\

\section{Summary}
In the present work, using an $R$-vine Markov-switching model, we aimed to find periods of \enquote{normal} and \enquote{abnormal} regimes within a data set consisting of North-American, European and Asian equity and volatility indices with an additional commodity index. We deliberately did not choose the usually applied wording of \enquote{non-crisis} and \enquote{crisis} as identified abnormality within the dependence structure does not necessarily go hand-in-hand with financial stress. After setting up sensible $R$-vine tree structures, we carried out an extensive rolling window analysis and drew the conclusion that regime changes mainly take place within the first trees. Estimating switching models for each continent individually and later on a joint global model shows the superiority of RV-MS setups over their corresponding static counterparty confirming the presence of global \enquote{normal} and \enquote{abnormal} regimes within the dependence structure of equity and implied volatility indices. In fact, the joint model performed best supporting the hypothesis of the existence of global regime switches.

\bibliographystyle{elsarticle-harv}
\bibliography{bib}
\end{document}